\documentclass[a4paper,reprint,nofootinbib,twoside,,showkeys]{revtex4-2}
\usepackage[left=20mm,right=20mm,top=25mm,columnsep=15pt]{geometry}
\usepackage[bookmarks=false,colorlinks]{hyperref}
\usepackage[english]{babel}
\usepackage[utf8]{inputenc}
\usepackage[T1]{fontenc}
\usepackage{amsmath,amsthm,amsfonts,amssymb,mathtools,bm,latexsym,stmaryrd}
\usepackage[table,dvipsnames]{xcolor}
\usepackage[pdftex]{graphicx}
\usepackage{adjustbox}
\usepackage{placeins}
\usepackage{lipsum}
\usepackage{csquotes}
\usepackage{comment}
\usepackage{subcaption}
\usepackage[title]{appendix}
\usepackage[bookmarks=false,colorlinks]{hyperref}
\hypersetup{urlcolor=NavyBlue, citecolor=NavyBlue}
\usepackage{ragged2e}
\usepackage{float}
\makeatletter
\let\newfloat\newfloat@ltx
\makeatother
\usepackage{algorithm}
\usepackage{algpseudocode}

\DeclareCaptionJustification{justified}{\justifying}
\DeclareMathOperator*{\argmax}{arg\,max}
\DeclareMathOperator*{\argmin}{arg\,min}
\def\eff{\text{eff}}
\def\opt{\text{opt}}

\begin{document}

\title{Routing algorithm within the multiple non-overlapping paths approach for quantum key distribution networks}

\author{Evgeniy O. Kiktenko}
\author{Andrey Tayduganov}\email[Correspondence email address:~]{a.taiduganov@misis.ru}
\author{Aleksey K. Fedorov}
\affiliation{National University of Science and Technology ``MISIS'', Moscow 119049, Russia}


\begin{abstract}
We develop a novel key routing algorithm for quantum key distribution (QKD) networks that utilizes a distribution of keys between remote, i.e., not directly connected by a QKD link, nodes through multiple non-overlapping paths.
This approach enchases the security of QKD network by minimizing potential vulnerabilities associated with individual trusted nodes. 
The algorithm ensures a balanced allocation of the workload across the QKD network links, while aiming for the target key generation rate between directly connected and remote nodes.
We present the results of testing the algorithm on two QKD network models consisting of 6 and 10 nodes. 
The testing demonstrates the ability of the algorithm to distribute secure keys among the nodes of the network in an all-to-all manner, ensuring that the information-theoretic security of the keys between remote nodes is maintained even when one of the trusted nodes is compromised.
These results highlight the potential of the algorithm to improve the performance of QKD networks.
\end{abstract}


\maketitle

\section{Introduction}

Quantum key distribution (QKD) technologies represent an elegant mechanism for solving the key distribution problem~\cite{Gisin2001,Scarani2009,Lo2014}, which is one of the central tasks of cryptography. 
Existing QKD protocols rely on the use of quantum channels for transmitting information that is encoded in quantum states of single photons
(on practice, the transmission of weak laser pulses via fiber-optic and free-space communication channels is used) and authentic classical channels~\cite{Gisin2001}.
The idea behind is that any interception in the process of transmitting quantum information can be detected by monitoring the quantum bit error rate (QBER) in the quantum channel, 
while authentication of classical communications is needed for the protection from man-in-the-middle attack during post-processing procedures~\cite{Gisin2001,Scarani2009}. 
In the view of the fact that the security of QKD is based on the laws of quantum physics, it is guaranteed to be secure against any unforeseen technological developments, such as cryptoanalysis using quantum computing~\cite{Shor1999}.
QKD technologies have attracted a great deal of interest and it is a widely studied field of quantum technologies~\cite{Lo2014,Diamanti2016}.
At the same time, existing realizations of QKD systems still face a number of challenges~\cite{Diamanti2016}, such as limitations in the key generation rate and distance, and also practical security. 

One of the ways to overcome the distance limitations, which is essentially related to losses during the transfer of quantum states, is to use QKD networks~\cite{Pan2018}. 
Moreover, QKD networks make key generation services available for multiple users. 
In the case of quantum communications, standard repeater schemes cannot be used, so quantum repeaters~\cite{Zoller1998,Lukin2001,Gisin2011} are required. 
Despite significant progress in this field during last decades (e.g., see Refs.~\cite{Lukin2020,Gisin2020,Lukin2024,Lukin20242}), quantum repeaters are not yet widely used in industrial QKD networks.

The most developed concept for large-scale, industrial QKD networks is to use trusted nodes~\cite{Elliott2004,Peev2009,Stucki2011,Pan2009,Pan2010,Han2010,Zeilinger2011}, 
which play a role of a user in QKD protocol that compose a common secret key by using the bit-wise XOR operation for two established keys with neighboring nodes.
The negative side for the trusted-node paradigm is that all the users of the network must trust the node and the node needs to be physically isolated well, 
while the positive side is reducing the cost and complexity as compared to the all connected point-to-point links.
At this stage, the deployment of QKD networks faces the problem of the optimal routing, which has a goal to optimally use the resources of the network when establishing keys between two arbitrary users. 
Existing QKD networks solve the routing problem by means of various protocols~\cite{Lutkenhaus2009,Tanizawa2016,Caleffi2017,Su2017,ma2017multiple,Mosca2018,Mehic2020,Amer2020,Chen2022,Chen2023,Bi2023,Dervisevic2024}, and tools (such as optical switches~\cite{Tang2006,Pan2010,Tayduganov2021}); 
however, this problem is far from being fully solved especially in the case of QKD networks of arbitrary topology~\cite{Mosca2018,Mehic2020,Amer2020,Chen2022,Chen2023,Bi2023,Dervisevic2024}.

In this work, we consider the key routing task within the multiple non-overlapping paths approach~\cite{salvail2010security,zhou2022quantum,gaidash2022quantum,solomons2022scalable,Stkepniak2023}.
The idea of this approach is to minimize the requirement for trust by distributing a key across remote nodes via independent routes. 
This ensures that the final key remains secure, provided that all trusted nodes on at least one route remain secure.
We propose a routing algorithm that is specifically designed for use with the multiple non-overlapping paths approach. This algorithm differs from previously considered algorithms for single-path routing~\cite{Lutkenhaus2009, Tanizawa2016, Su2017,Chen2022, Bi2023} in that it is tailored to handle the unique requirements of the multiple non-overlapping path scenario.
The developed algorithm differs from that proposed in~\cite{ma2017multiple} by incorporating the limited capacity for key generation in networks and addressing the routing issue for all nodes simultaneously. At the same time, it differs from the more recent scheme for hybrid-trusted networks~\cite{Chen2023} in that all nodes in the network are assumed to have equal trust in it.
Importantly, key routing algorithms designed for the standard single-path trusted node schemes do not guarantee the efficiency of key distribution in the context of the non-overlapping paths approach. 
To address this challenge, we propose and test the first, to our knowledge, key routing algorithm specifically tailored for the non-overlapping paths framework.

The paper is organized as follows.
In Sec.~\ref{sec:M-path}, we discuss the basic features of the multiple non-overlapping paths approach.
In Sec.~\ref{sec:Algorithm}, we present the developed routing algorithm.
In Sec.~\ref{sec:Performance}, we demonstrate the performance of the algorithm on two QKD network models.
Finally, we conclude in Sec.~\ref{sec:Conclusion}.

\section{The concept of the $M$--paths scheme}\label{sec:M-path}

To overcome the limitation of the QKD range, multiple short point-to-point QKD links are combined into QKD networks with so-called intermediate trusted nodes~\cite{Pan2018,Pan2021}. 
In this approach, quantum keys are generated only between the network nodes connected directly by a quantum channel, and trusted nodes perform a key hopping via subsequent bit-wise XOR operations. 
With the use of this technology, it is possible to build QKD networks of various topologies like backbone, ring, star or mixed one.

The downside of the trusted node paradigm is that remote network nodes have to trust the physical integrity of all nodes in the chain between them. 
It is possible to relax the trust requirement and increase the security by considering an $M$ non-overlapping paths approach~\cite{salvail2010security,zhou2022quantum,gaidash2022quantum,solomons2022scalable,Stkepniak2023}, 
where the secret key between two remote nodes is obtained by combining keys from $M$ (with $M>1$) \emph{non-overlapping} paths.
In the following, we refer to it as $M$--paths approach.
We would like to note that, in this context, we refer to a path as an independent key hopping route. 
From a physical perspective, such independent paths can be realized even within a single transmission line, where certain nodes may be skipped, as is the case of quantum control-based key distribution networks \cite{Kirsanov2023,kirsanov2024loss}.

Let us consider a communication between remote nodes $i$ and $j$ via a set of $M$ non-overlapping paths ${\cal P}_{ij}^M$. 
Let each path $P\in{\cal P}_{ij}^M$ provide an $\ell$--bit information-theoretically secure key ${\cal K}_{ij}^P$. 
Then the final $\ell$--bit secret key to be used for the unconditionally secure data encryption with the one-time pad (OTP) technique can be obtained as follows:
\begin{equation}
    {\cal K}_{ij} = \bigoplus_{P\in{\cal P}_{ij}^M} {\cal K}_{ij}^P \,,
\end{equation}
where $\bigoplus$ stands for a bit-wise exclusive or (XOR) operation. 
Therefore, due to the perfect secrecy of the OTP, the eavesdropping becomes technically more complicated in the $M$--path scheme since the intermediate nodes from each of $M$ paths must be hacked (compromised) in order to reconstruct the entire secret key.
Equivalently, compromising fewer than $M$ nodes from ${\cal P}_{ij}^M$ would not compromise ${\cal K}_{ij}$.
Assuming a probability of compromise of a node operated as a trusted node is $\epsilon_{\rm compr}$, we can estimate that the probability of compromising a key between any two remote nodes in $M$--paths regime is upper bounded by $\epsilon^M_{\rm compr}$.

Here, we would like to emphasize several points related to the $M$--paths scheme.
First, the applicability of $M$--path schemes strictly depends on the topology of QKD networks. 
It is impossible to use this scheme in backbone-type networks, where there is only a single path between two remote nodes, or to apply a 3--path scheme within a ring topology of the network graph.
Second, distributing an $\ell$--bit key using the $M$--path scheme reduces the remaining keys by $\ell$ bits for each direct link in each path (we discuss this in more detail in the following section). 
Therefore, increasing $M$ also increases the ``cost'' of distributing keys between remote nodes in terms of the remaining key between network nodes.
In this regard, the selection of $M$ within the range allowed by the topology should consider a balance between the desired security level, estimated as $\epsilon_{\rm compr}^M$, and the increased key usage in the network.
Thirdly, the secret key ${\cal K}_{ij}$ can be obtained by concatenating several keys corresponding to different $M$--element sets of non-overlapping paths $\{{\cal P}_{ij}^{M(k)}\}_k$:
\begin{equation}
    {\cal K}_{ij} = \bigg(\bigoplus_{P\in{\cal P}_{ij}^{M(1)}} {\cal K}_{ij}^P\bigg) \big\| \bigg(\bigoplus_{P'\in{\cal P}_{ij}^{M(2)}} {\cal K}_{ij}^{P'}\bigg) \big\| \ldots\,,
    \label{eq:several_M-paths}
\end{equation}
where $\|$ denotes concatenation.
It should be noted that some of the paths in the sets $\{{\cal P}_{ij}^{M(k)}\}_k$  may have common elements.
Additionally, the lengths of concatenated parts may differ.
The construction of ${\cal K}_{ij}$ in the form of Eq.~\eqref{eq:several_M-paths}, assuming it is permitted by the network topology, enables a more balanced distribution of key consumption across the network.
The last point raises an important challenge in finding a practical and efficient key routing algorithm within the $M$--path approach for QKD networks of arbitrary topology. 
This algorithm should allow for the selection of suitable combinations of paths in order to distribute keys between pairs of remote nodes. 
Additionally, it should ensure a sufficient reserve of available keys for directly connected nodes.
We consider the construction of such an algorithm in the section below.

\section{Routing algorithm for the $M$--path scheme}\label{sec:Algorithm}

We begin with introducing some formal definitions and notations used later.
An arbitrary QKD network with $N$ nodes is represented as a graph $G=(V,E)$, where the vertices $V=\{i\}_{i=0}^{N-1}$ are the network nodes and the edges $E\subseteq\{(i,j)|\,i,j\in V\}$ are the fiber optic lines connecting them. 
The edge weight corresponds to the secret key generation rate of the corresponding section of the fiber optic line. 
The definition of the $M$--path routing scheme obviously implies that the degree of each vertex, i.e. the number of edges that are incident to the vertex, 
is required to be $\deg(i)\geq M$. A path in a graph is a finite sequence of edges which joins a sequence of distinct vertices. 
An open path from node $i$ to node $j$ can be defined in a short way as a sequence of adjacent vertices, connected by edges,
\begin{equation}
    P_{ij} \coloneqq (i,k,l,\dots,m,j) \,,
\end{equation}
with $(i,k),(k,l),\dots,(m,j)\in E$.

Let $R_{ij}=R_{ji}$ be the secret key generation rate for a pairwise QKD link or chain of links connecting nodes $i$ and $j$ such that $R_{ij}>0$ if $(i,j)\in E$ and is zero, otherwise. 
The average length of the local secret key, $K_{ij}$, accumulated during some time period $\tau$, is equal to $|K_{ij}|=R_{ij}\tau$.
Let $T_{ij}$ be a target secret key generation rate between the nodes $i$ and $j$, which can be associated with the key consumption rate.
Note that it is possible to have $T_{ij}>0$ even if $(i,j)\not\in E$.
Then the goal is to develop a (sub)optimal key flow that provides the desired $T_{ij}$ for all pairs $(i,j)$ by manipulating the effective key rates $R_{ij}^\eff$.

\begin{figure}[t!]\centering
	\includegraphics[width=0.99\linewidth]{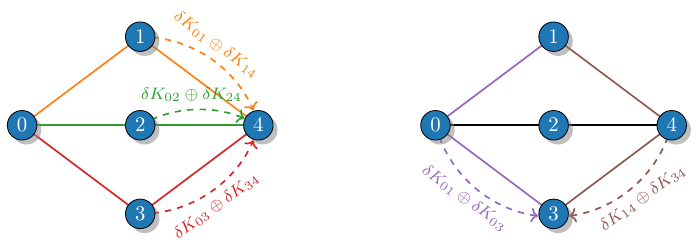}
    \captionsetup{justification=justified}
	\caption{The key distribution between the remote nodes 0 and 4 (left) and 1 and 3 (right) via various non-overlapping paths for a network of five nodes. The path index of $\delta K_{ij}$ is omitted for simplicity.}
	\label{fig:network_example}
\end{figure}

To illustrate the concept of the effective rate, consider a simple network of five nodes in Fig.~\ref{fig:network_example} and set $M=2$. 
The communication between the remote nodes 0 and 4 ($R_{04}\equiv0$) in the $2$--path scheme can be done via three possible pairs of non-overlapping paths,
\begin{equation}
    \begin{split}
        \mathcal{P}_{04}^{2(1)} &= \{(0,1,4),(0,2,4)\} \,, \\
        \mathcal{P}_{04}^{2(2)} &= \{(0,2,4),(0,3,4)\} \,, \\
        \mathcal{P}_{04}^{2(3)} &= \{(0,1,4),(0,3,4)\} \,,
        \label{eq:04_path_pairs}
    \end{split}
\end{equation}
while the communication between the nodes 1 and 3 ($R_{13}\equiv0$) is feasible only via one pair,
\begin{equation}
    \mathcal{P}_{13}^{2} = \{(1,0,3),(1,4,3)\} \,.
\end{equation}
Every local accumulated secret key $K_{ij}$ is split into subsamples in the following way:
\begin{equation}
    \begin{split}
        K_{01} &= K_{01}^\eff \| \delta K_{01}^{(0,1,4)} \| \delta K_{01}^{(0,1,4)\prime} \| \delta K_{01}^{(1,0,3)} \,, \\
        K_{02} &= K_{02}^\eff \| \delta K_{02}^{(0,2,4)} \| \delta K_{02}^{(0,2,4)\prime} \,, \\
        K_{03} &= K_{03}^\eff \| \delta K_{03}^{(0,3,4)} \| \delta K_{03}^{(0,3,4)\prime} \| \delta K_{03}^{(1,0,3)} \,, \\
        K_{14} &= K_{14}^\eff \| \delta K_{14}^{(0,1,4)} \| \delta K_{14}^{(0,1,4)\prime} \| \delta K_{14}^{(1,4,3)} \,, \\
        K_{24} &= K_{24}^\eff \| \delta K_{24}^{(0,2,4)} \| \delta K_{24}^{(0,2,4)\prime} \,, \\
        K_{34} &= K_{34}^\eff \| \delta K_{34}^{(0,3,4)} \| \delta K_{34}^{(0,3,4)\prime} \| \delta K_{34}^{(1,4,3)} \,,
    \end{split}
    \label{eq:key_split_5v}
\end{equation}
where the keys $\delta K_{ij}^{P(\prime)}$ are sacrificed for the key hopping through the path $P$ with OTP technique, while the remaining effective keys $K_{ij}^\eff$ are used for direct communication between connected nodes. 
We note that $\delta K_{ij}^P\neq \delta K_{ij}^{P\prime}$ since the path belongs to different $M$--path sets, see Eq.~\eqref{eq:04_path_pairs}. 
The OTP technique requires the key lengths to satisfy the following conditions:
\begin{equation*}
    \begin{split}
        & |\delta K_{01}^{(0,1,4)}| = |\delta K_{14}^{(0,1,4)}| = |\delta K_{02}^{(0,2,4)}| = |\delta K_{24}^{(0,2,4)}| \,, \\
        & |\delta K_{02}^{(0,2,4)\prime}| = |\delta K_{24}^{(0,2,4)\prime}| = |\delta K_{03}^{(0,3,4)}| = |\delta K_{34}^{(0,3,4)}| \,, \\
        & |\delta K_{01}^{(0,1,4)\prime}| = |\delta K_{14}^{(0,1,4)\prime}| = |\delta K_{03}^{(0,3,4)\prime}| = |\delta K_{34}^{(0,3,4)\prime}| \,, \\
        & |\delta K_{01}^{(1,0,3)}| = |\delta K_{03}^{(1,0,3)}| = |\delta K_{14}^{(1,4,3)}| = |\delta K_{34}^{(1,4,3)}| \,.
    \end{split}
\end{equation*}
Then the final effective keys to encrypt the communication between remote nodes are formed according to Eq.~\eqref{eq:several_M-paths},
\begin{equation}
    \begin{split}
        K_{04}^\eff &=
            \big(\delta K_{01}^{(0,1,4)} \oplus \delta K_{02}^{(0,2,4)}\big) \\
            &~~\,\|\big(\delta K_{02}^{(0,2,4)\prime} \oplus \delta K_{03}^{(0,3,4)}\big) \\
            &~~\,\|\big(\delta K_{01}^{(0,1,4)\prime} \oplus \delta K_{03}^{(0,3,4)\prime}\big) \,, \\
        K_{13}^\eff &= \delta K_{01}^{(1,0,3)} \oplus \delta K_{14}^{(1,4,3)} \,.
    \end{split}
\end{equation}
Now having all nodes ``connected'', we can define the effective generation rate for arbitrary pair of nodes in the network as
\begin{equation}
    R_{ij}^\eff \coloneqq \frac{|K_{ij}^\eff|}{\tau} > 0 \,. 
\end{equation}
It is easy to see that $R_{ij}^\eff<R_{ij}$ for all $(i,j)\in E$. 
In this way, the key management task is to find an optimal set of $\delta K_{ij}^{P(\prime)}$ in Eq.~\eqref{eq:key_split_5v} such that $R_{ij}^\eff\geq T_{ij}$ for all $i,j\in V$.

Here, we would like to draw attention to the fact  that all OTP-secured communications must be authenticated, which also requires the use of a secret key.
Due to the fixed consumption of secret keys in authentication of each message, it is more efficient in practice to increase the interval between ``key distribution rounds'' and, consequently, the length of the keys transmitted in each message. 
This approach helps to minimize the relative costs associated with authentication.
In the following, we assume that the key consumption for authentication can be neglected compared to the transmitted key material.

The proposed routing Algorithm~\ref{alg} follows an iterative process (see also an example of a step-by-step workflow in Appendix~\ref{app}). 
As the basic target quantity to be minimized in each iteration we consider
\begin{equation} \label{eq:costfunc}
	\Delta \coloneqq \max_{i,j\in V}\big[T_{ij}-R_{ij}^\eff\big] \,,
\end{equation}
which provides the discrepancy between the target and effective rates for the ``worst'' pair of nodes. 
Having $\Delta\leq0$ means that the goal is achieved, and the algorithm can be stopped.
At every iteration $r=1,2,\dots,r_{\max}$, where the hyperparameter $r_{\max}$ stands for the maximal allowed number of iterations, first we search for the currently ``worst'' pair of nodes in the graph (i.e. with maximum value of $\Delta$), denoted by $(i^*,j^*)$. 
If the found pair is connected directly, then the algorithm stops.
In the case of multiple pair candidates a random candidate is selected.
Then among all possible sets of $M$ non-overlapping paths from $i^*$ to $j^*$, 
denoted by ${\cal L}_{i^*j^*}^M$ the optimal set ${\cal P}_{i^*j^*}^{M,\opt}$ is selected by searching for a set that has a path containing a pair of connected nodes with the minimal rate deficiency. 
In~Algorithm~\ref{alg}, we formalize it by introducing the deficiency function ${\bf D}(\cdot)$, which returns the maximal deficiency among all direct links within input set of $M$ non-overlapping paths.
Again, if multiple candidates are found, a random candidate over candidates with the minimal total length is selected (this is formalized by $\texttt{rand\textunderscore ch\textunderscore min\textunderscore dist}(\cdot)$ function).
When the optimal $M$--path combination is determined, $R_{i^*j^*}^{\eff}$ is increased by the small amount $\delta R$ (comes as an input hyperparameter of the algorithm), 
while $R_{ij}^{\eff}$ is decreased by $\delta R$ for all adjacent nodes from every path in ${\cal P}_{i^*j^*}^{M,\opt}$.

After that, we calculate the new value of the cost function according to Eq.~\ref{eq:costfunc} and store it in $\Delta_{\rm upd}$.
If the new value is ``worse'' than the one in the begining of the iteration ($\Delta_{\rm upd}>\Delta$), then the algorithm stops, otherwise the information about obtained key routing way ${\cal P}_{i^*j^*}^{M,\opt}$ between $i^*$ and $j^*$ is added to the list ${\sf routing\_list}$, 
and the algorithm proceeds to the next iteration.
Here we assume that ${\sf routing\_list}$ consists of records of the form $({\cal P}_{i'j'}^{M}:R')$.
Each of such records means that during some fixed time period $\tau$, a key of length $R'\tau$ have to be distributed between $i'$ and $j'$ over $M$ paths in ${\cal P}_{i'j'}^{M}$.
Thus, if ${\sf routing\_list}$ already contains a record of the form $({\cal P}_{i^*j^*}^{M,\opt}:R^{\rm cur}_{i^*j^*})$, then it is replaced by $({\cal P}_{i^*j^*}^{M,\opt}:R^{\rm cur}_{i^*j^*}+\delta R)$, or the new record $({\cal P}_{i^*j^*}^{M,\opt}:\delta R)$ is added, otherwise.

\begin{algorithm}[t!]
    \begin{algorithmic}
        \Require $(R_{ij})$, $(T_{ij})$, $\delta R$, $r_{\max}$
        \Ensure {\sf routing\_list}
        \State $R_{ij}^{\eff}=R_{ij}~~\forall i,j\in V$ \Comment{initialization of the effective key generation matrix}
        \State $r=0$ \Comment{initialization of iteration counter}
        \State {\sf routing\_list} = [\,] \Comment{initialization of empty routing list}
        \State $\Delta=\max_{i,j\in V}[T_{ij}-R_{ij}]$
        \While{$\Delta>0\,\wedge\,r<r_{\max}$}
            \State $D(i,j)\coloneqq T_{ij}-R_{ij}^{\eff}$ \Comment{current rate deficiency function}
            \State $(i^*,j^*)=\texttt{rand\textunderscore ch}\{\argmax_{i,j\in V} D(i,j)\}$ \Comment{selecting a random pair over pairs with the ``worst'' deficiency}
            \If{$(i^*,j^*)\in E$}
                \State {\bf return} {\sf routing\_list} \Comment{stopping the algorithm}
            \EndIf
            \State ${\cal P}_{i^*j^*}=\{P=(i^*,\dots,j^*)\}$ \Comment{set of all possible paths from $i^*$ to $j^*$}
            \State ${\cal L}^{M}_{i^*j^*}=\{{\cal P}_{i^*j^*}^M=\{P\in{\cal P}_{i^*j^*}:P\cap P^\prime=\{i^*,j^*\}~\forall P^\prime\in{\cal P}_{i^*j^*}^M, P^\prime\neq P\}:|{\cal P}_{i^*j^*}^M|=M\}$ \Comment{list of all possible combinations of $M$ non-overlapping paths from $i^*$ to $j^*$}
            \State ${\bf D}({\cal P}_{i^*j^*}^M)\coloneqq\max_{P\in{\cal P}_{i^*j^*}^M}\max_{(i,j)\in P}D(i,j)$ \Comment{deficiency function which characterizes the $M$--path combination capacity (``deficiency of the worst link in the worst path'')}
            \State ${\cal P}_{i^*j^*}^{M,\text{pre-opt}} =\argmin_{{\cal P}_{i^*j^*}^M\in{\cal L}}{\bf D}({\cal P}_{i^*j^*}^M)$
            \State ${\cal P}_{i^*j^*}^{M,\text{opt}}=\texttt{rand\textunderscore ch\textunderscore min\textunderscore dist}\{{\cal P}_{i^*j^*}^{M,\text{pre-opt}} \}$ \Comment{taking the optimal $M$--path combination}
            \State $R_{i^*j^*}^{\eff}=R_{i^*j^*}^{\eff}+\delta R$ \Comment{increase the effective key rate for the ``worst'' pair}
            \ForAll{$P\in{\cal P}_{i^*j^*}^{M,\text{opt}}$}
                \ForAll{$i,j\in P:(i,j)\in E$}
                    \State $R_{ij}^{\eff}=R_{ij}^{\eff}-\delta R$ \Comment{decrease the effective key rates for adjacent pairs}
                \EndFor
            \EndFor
            \State $\Delta_{\rm upd}=\max_{i,j\in V}[T_{ij}-R_{ij}^\eff]$
            \If{$\Delta_{\rm upd} \leq \Delta$}
                \State update ${\sf routing\_list}$ with a record $({\cal P}_{i^*j^*}^{M,\text{opt}}: \delta R)$
                \State $\Delta=\Delta_{\rm upd}$ \Comment{updating the value if the cost and function}
                \State $r=r+1$ \Comment{incrementing the counter}
            \Else 
                \State {\bf return} {\sf routing\_list} \Comment{stopping the algorithm}
            \EndIf
        \EndWhile
        \State {\bf return} {\sf routing\_list}
    \end{algorithmic}
    \caption{}
    \label{alg}
\end{algorithm}

Finally, we would like to discuss the role of two hyperparameters: $r_{\max}$ and $\delta R$. 
By bounding the number of iterations, $r_{\max}$ limits the maximum running time of the algorithm. 
One can also consider replacing the condition $r < r_{\max}$ in the main loop of the algorithm by a condition that checks if the current running time is less than the maximum allowed time. If there are no limits on running time, $r_{\max}$ can be set to infinity.
$\delta R$ determines the size of the step at each iteration. A smaller $\delta R$ results in more iterations and a higher precision in the final value of the cost function ($\Delta$).
As we will see in the testing in the next section, reducing $\delta R$ can improve the cost function by the amount of $\delta R$. 
However, from a practical perspective, it is not advisable to reduce $\delta R$ below a level corresponding to the typical cost of a secret key for authentication purposes, assumed to be negligible within the construction of the algorithm.

\section{Performance demonstration of the algorithm}\label{sec:Performance}

To demonstrate the performance of the developed algorithm, we consider the task of routing the key within the $2$--path scheme for two examples of QKD networks.

\begin{figure}[t!]\centering
    \begin{subfigure}[t!]{0.5\linewidth}\centering
        \includegraphics[width=0.7\linewidth]{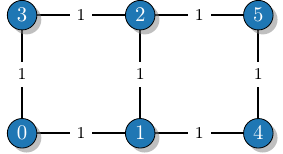}\\
        \caption{}
        \label{fig:network_6v_graph}
        \includegraphics[width=0.99\linewidth]{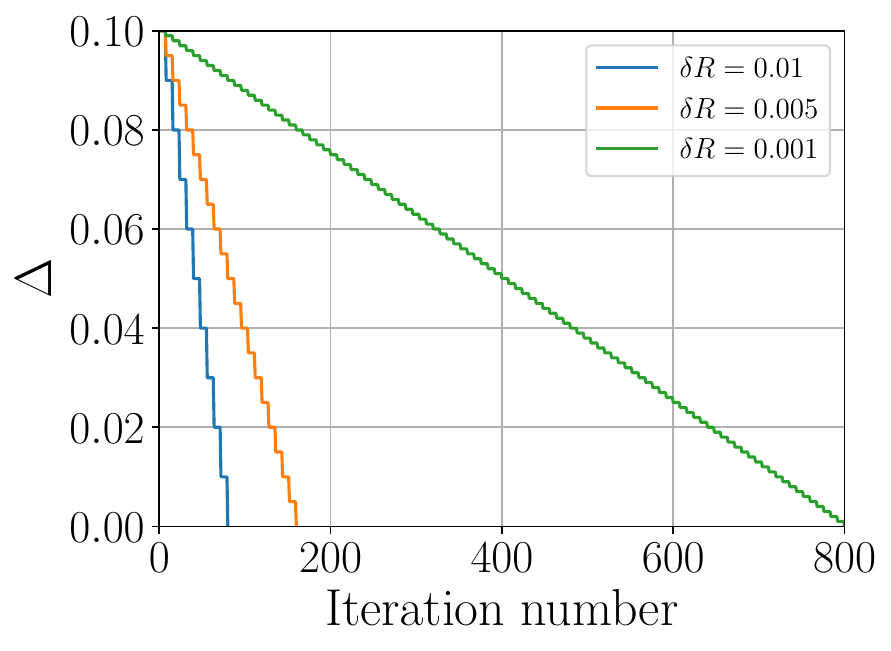}
        \caption{}
        \label{fig:network_6v_cost}
    \end{subfigure}
    \begin{subfigure}[t!]{0.45\linewidth}\centering
        \includegraphics[width=0.99\linewidth]{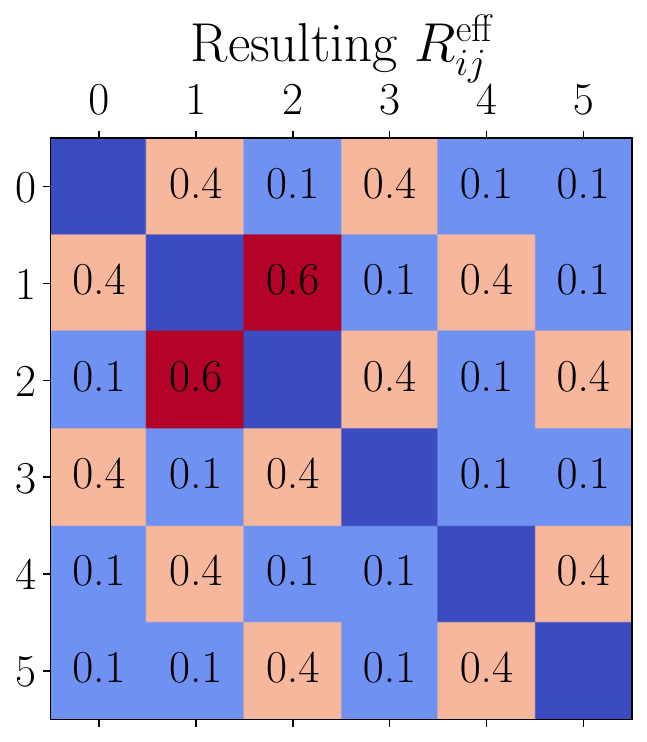}
        \caption{}
        \label{fig:network_6v_matrix}
    \end{subfigure}
    \caption{(a) The network graph with edges, labeled with secret key generation rates in kbps. (b) The iteration number dependence of the cost function. (c) The resulting effective secret key generation rates for various pairs of network nodes.}
    \label{fig:network_6v}
\end{figure}

The first QKD network is represented by a six-vertex graph shown in Fig.~\ref{fig:network_6v_graph}.
For simplicity, we assume that all adjacent links are equidistantly located and have equal initial key generation rates, namely, $R_{ij}\coloneqq1$\,kbit/s for all $(i,j)\in E$. The target rates are also set to be equal, $T_{ij}\coloneqq0.1$\,kbit/s, for all possible pairs. 
The decrease of the cost function $\Delta$ with iteration number $r$  is shown in Fig.~\ref{fig:network_6v_cost}. The algorithm completes the task after 80, 160 and 800 iterations, respectively. For all considered values of $\delta R$, the matrix of final effective rates  $R^\eff$ approaches to a form shown in Fig.~\ref{fig:network_6v_matrix}. The corresponding output routing list for remote nodes is obtained in the following:
\begin{equation}
    \begin{aligned}
        {\sf routing\_list} = [
        \{(0, 1, 2), (0, 3, 2)\} &: 0.1\,,\\
        \{(1, 0, 3), (1, 2, 3)\} &: 0.1\,,\\
        \{(1, 2, 5), (1, 4, 5)\} &: 0.1\,,\\
        \{(2, 1, 4), (2, 5, 4)\} &: 0.1\,,\\
        \{(0, 1, 4, 5), (0, 3, 2, 5)\} &: 0.1\,,\\
        \{(3, 0, 1, 4), (3, 2, 5, 4)\} &: 0.1\,,\\
        \{(0, 1, 4), (0, 3, 2, 5, 4)\} &: 0.1\,, \\
        \{(3, 0, 1, 4, 5), (3, 2, 5)\} &: 0.1] \,.\\
    \end{aligned}
    \label{eq:paths_6v}
\end{equation}

It can be seen that the resulting paths follow the graph symmetry and match intuitive expectations.
It should be noted that each pair of remote nodes $(i,j)$ has a unique pair of paths through which the key is distributed between them. Considering the key management, the accumulated secret key, e.g. $K_{01}$, is split into multiple subsamples: six keys $\{\delta K_{01}^{P}\}$ of equal length $|\delta K_{01}^P|=0.1|K_{01}|$ are used for the key hopping via the paths (0,1,2), (1,0,3), (0,1,4,5), (3,0,1,4), (0,1,4) and (3,0,1,4,5), while the remaining key $K_{01}^\eff=K_{01}\backslash\left(\bigcup_P\delta K_{01}^P\right)$ of length $|K_{01}^\eff|=0.4|K_{01}|$ can be entirely used for the direct communication between the nodes 0 and 1. A similar key splitting procedure according to the $R^\eff$ matrix in Fig.~\ref{fig:network_6v_matrix} is performed for other edges of the graph. Then the effective secret key to encrypt the communication between a pair of remote nodes, e.g. 1 and 3, is $K_{13}^\eff=\delta K_{01}^{(1,0,3)}\oplus\delta K_{12}^{(1,2,3)}$.

\begin{figure}[t!]\centering
    \begin{subfigure}[t!]{0.5\linewidth}\centering
        \includegraphics[width=0.99\linewidth]{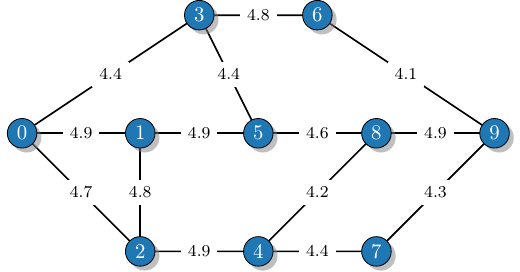}\\
        \caption{}
        \label{fig:network_10v_graph}
        \includegraphics[width=0.99\linewidth]{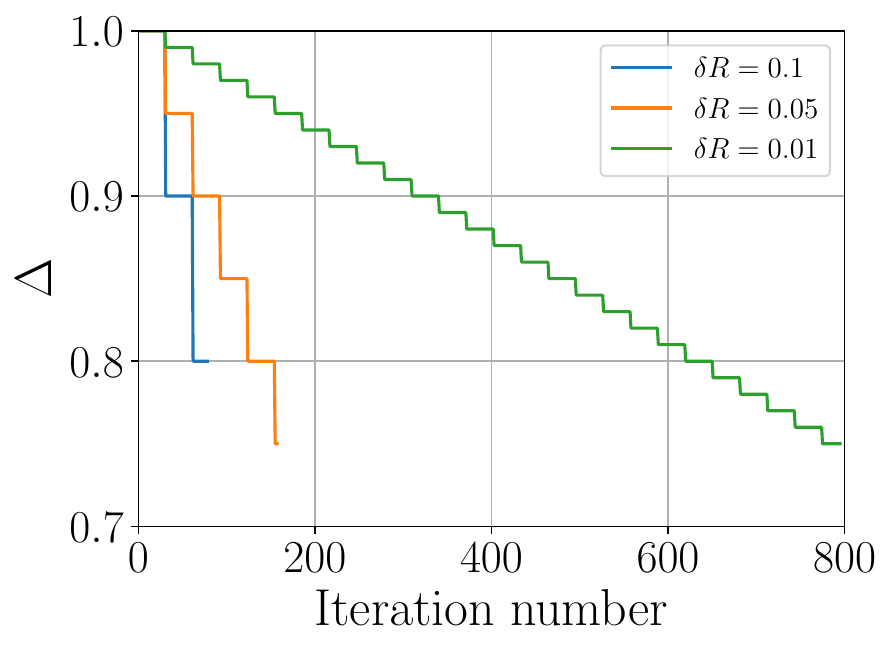}
        \caption{}
        \label{fig:network_10v_cost}
    \end{subfigure}
    \begin{subfigure}[t!]{0.45\linewidth}\centering
        \includegraphics[width=0.99\linewidth]{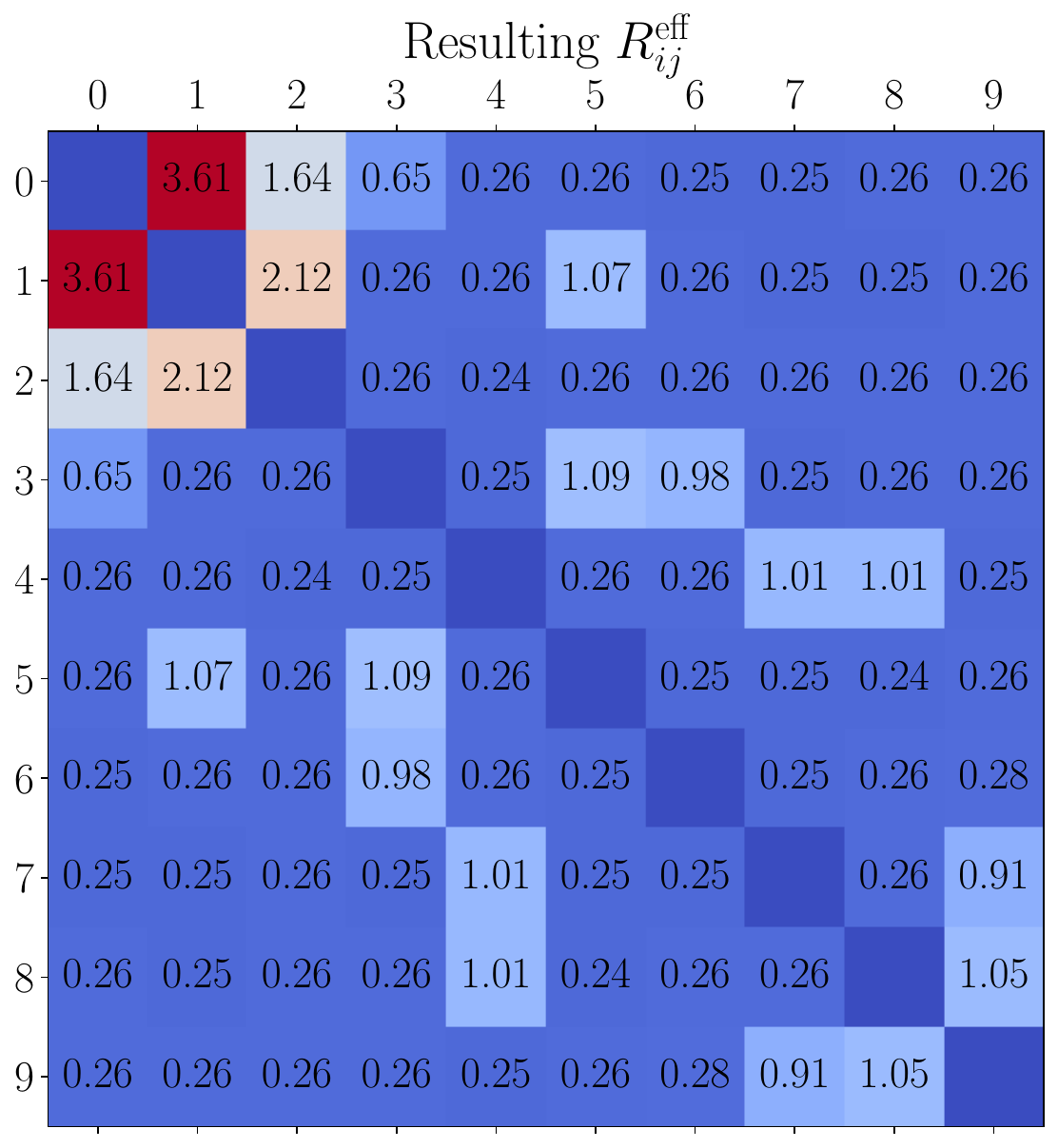}
        \caption{}
        \label{fig:network_10v_matrix}
    \end{subfigure}
    \caption{(a) The network graph from Ref.~\protect\cite{Chen2022} with edges that are labeled with the secret key generation rates in kbps, simulated for the decoy-state BB84 protocol using the simple theoretical model from Ref.~\protect\cite{Ma2005} and taking into account the detector's dead time effect. 
    (b) The iteration number dependence of the cost function. 
    (c) The resulting effective secret key generation rates (obtained with $\delta R=0.01$) for various pairs of network nodes.}
    \label{fig:network_10v}
\end{figure}

The second considered example of a more complex QKD network topology, which is taken from Ref.~\cite{Chen2022}, is shown in Fig.~\ref{fig:network_10v_graph}.
In contrast to the previous case, the network has different rates for each link and does not possess a symmetrical structure.
The secret key generation rates are simulated for the decoy-state BB84 protocol~\cite{Wang2005,Ma2005,Zhang2017,Trushechkin2017,Trushechkin2021} using the link distances from the corresponding graph in Ref.~\cite{Chen2022}.
For this example, we set $T_{ij}\coloneqq1.0$\,kbit/s for all possible pairs of nodes and $M=2$. 
The behavior of the cost function $\Delta$ with iteration number $r$ for $\delta R\in\{0.1,0.05,0.01\}$ is shown in Fig.~\ref{fig:network_10v_cost}.
The algorithm stops after 78, 158, and 793 iterations, respectively.
One can notice that $\Delta$ does not approach to zero like in the first example and stops around 0.75 since the chosen target matrix turns out to be overrated for this scheme.
We also note the choice of $\delta R$ slightly affects the resulting value of the cost function.

The resulting form of the $R^{\rm eff}$ matrix for $\delta R=0.01$, shown in Fig.~\ref{fig:network_10v_matrix}, demonstrates that the algorithm provide close-to-uniform distribution of the effective key generation rates among remote nodes.
The 11 links, which keep an effective key generation rate noticeably above a ``water level'' of 0.25, correspond to directly connected nodes. 
However, note that the key generation in direct links $(2,4)$, $(5,8)$, and $(6,9)$ drops to the ``water level'', stopping the algorithm.

For additional illustration, we present all the path pairs from {\sf routing\_list} used for the key distribution between 0th and 9th nodes:
\begin{equation}
    \begin{aligned} 
        {(0, 2, 4, 7, 9), (0, 3, 5, 8, 9)} &: 0.02\,,\\
        {(0, 1, 5, 8, 9), (0, 2, 4, 7, 9)} &: 0.03\,,\\
        {(0, 1, 5, 8, 9), (0, 3, 6, 9)} &: 0.1\,,\\
        {(0, 2, 4, 7, 9), (0, 3, 6, 9)} &: 0.1\,,\\
        {(0, 2, 4, 8, 9), (0, 3, 6, 9)} &: 0.01 \,.
    \end{aligned}
\end{equation}
Notably, the effective key generation rate of $R_{09}^{\rm eff}=0.26$ is realized in a quite non-trivial way via five different non-overlapping path pairs.

Then the final effective secret key between the nodes 0 and 9, assembled according to Eq.~\eqref{eq:several_M-paths}, can be written as
\begin{equation}
    \begin{split}
        K_{09}^\eff =
        &\big(\delta K_{02}^{(0,2,4,7,9)} \oplus \delta K_{03}^{(0,3,5,8,9)}\big) \\
        \|&\big(\delta K_{01}^{(0,1,5,8,9)} \oplus \delta K_{02}^{(0,2,4,7,9)\prime}\big) \\
        \|&\big(\delta K_{01}^{(0,1,5,8,9)\prime} \oplus \delta K_{03}^{(0,3,6,9)}\big) \\
        \|&\big(\delta K_{02}^{(0,2,4,7,9)\prime\prime} \oplus \delta K_{03}^{(0,3,6,9)\prime}\big) \\
        \|&\big(\delta K_{02}^{(0,2,4,8,9)} \oplus \delta K_{03}^{(0,3,6,9)\prime\prime}\big) \,,
    \end{split}
\end{equation}
where the lengths of the XORed pairs of keys depend on the respective effective rates. A more detailed description of such key management becomes highly non-trivial in this case and is beyond this research.

\section{Conclusion and outlook}\label{sec:Conclusion}

In this work, we have addressed the issue of trust reduction in QKD networks by studying the key distribution among remote nodes through multiple non-overlapping paths ($M$--path scheme). 
We have proposed a novel iterative greedy algorithm for key routing in QKD networks, aimed at effectively facilitating the key distribution between remote nodes while utilizing the $M$--path scheme. 
The efficiency of the proposed algorithm has been shown through case studies involving two QKD networks comprising 6 and 10 nodes, respectively.
Our results have demonstrated the potential for generating non-trivial key routing paths between remote nodes while achieving balanced load distribution across directly connected network nodes. 
This work not only contributes to enhancing the reliability of QKD systems, but also opens avenues for further research in optimizing key distribution strategies for QKD networks.

As a potential limitation on the applicability of the developed algorithm, we would like to emphasize its strong dependence on network topology. 
Specifically, it is assumed that any two remote nodes can be connected by at least $M$ non-overlapping paths. 
However, in a real QKD network, this assumption may not hold. It is possible to envision a scenario where some pairs of remote nodes are connected by non-overlapping paths, but others are not. 
This limitation can be addressed by considering a more flexible approach with different requirements for the value of $M$ for each pair, or by considering ``partially overlapping path'' scenarios. 
This appears to be a promising avenue for future research.

\acknowledgments{This research was supported by the Priority 2030 program at the National University of Science and Technology ``MISIS'' under the project K1-2022-027. 
We also thank V. Dubinin for fruitful discussions.}


\begin{appendices}

\section[\appendixname~\thesection]{}\label{app}

\begin{figure}[t!]\centering
    \begin{subfigure}[t!]{0.45\linewidth}\centering
        \includegraphics[width=0.99\linewidth]{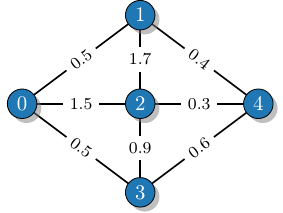}\\
        \caption{}
        \label{fig:network_5v_graph}
    \end{subfigure}
    \begin{subfigure}[t!]{0.45\linewidth}\centering
        \includegraphics[width=0.99\linewidth]{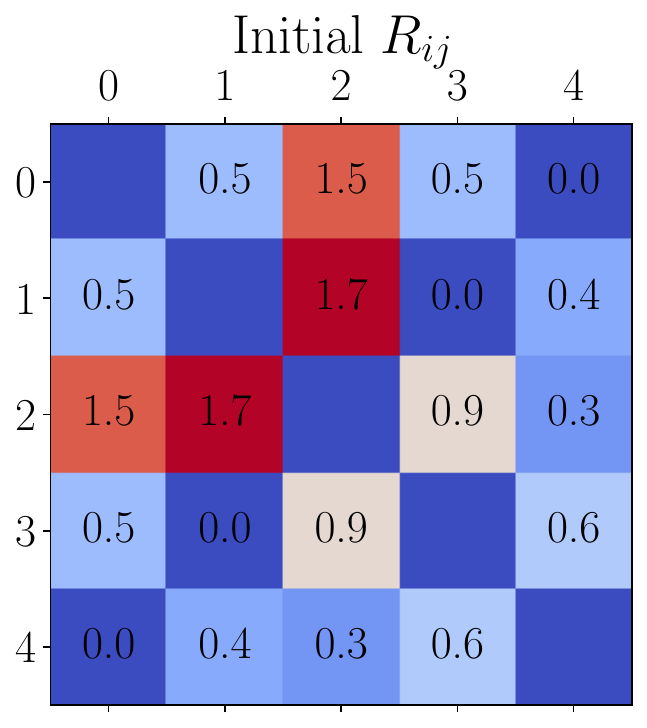}
        \caption{}
        \label{fig:network_5v_matrix_in}
    \end{subfigure}
    \begin{subfigure}[t!]{0.45\linewidth}\centering
        \includegraphics[width=0.99\linewidth]{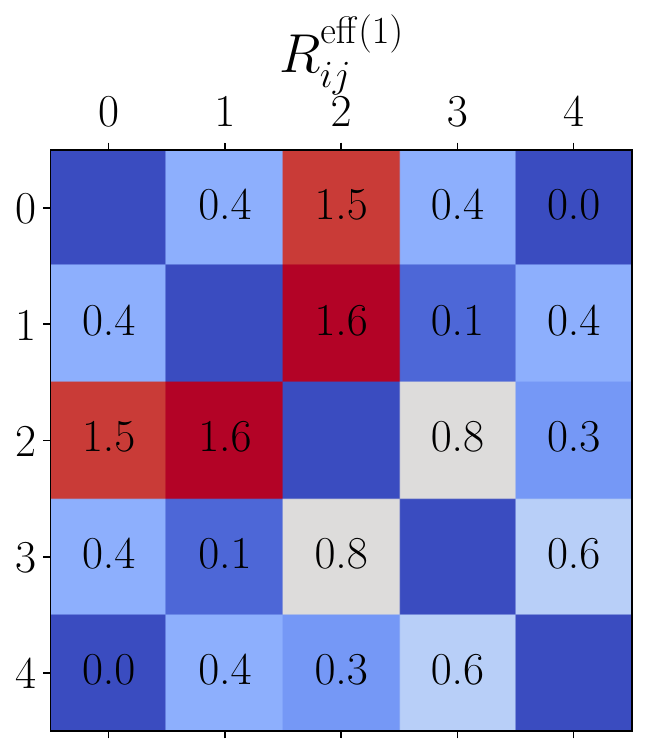}
        \caption{}
        \label{fig:network_5v_iter_1}
    \end{subfigure}
    \begin{subfigure}[t!]{0.45\linewidth}\centering
        \includegraphics[width=0.99\linewidth]{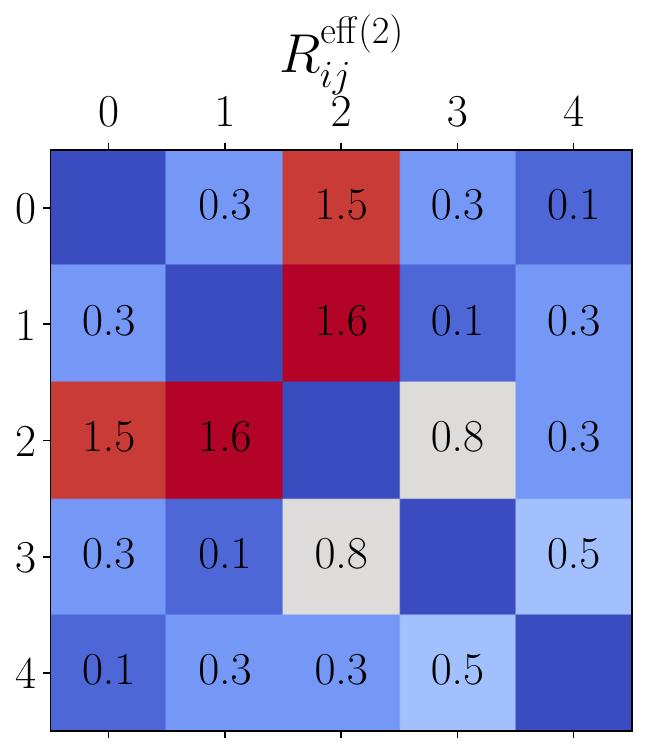}
        \caption{}
        \label{fig:network_5v_iter_2}
    \end{subfigure}
    \begin{subfigure}[t!]{0.45\linewidth}\centering
        \includegraphics[width=0.99\linewidth]{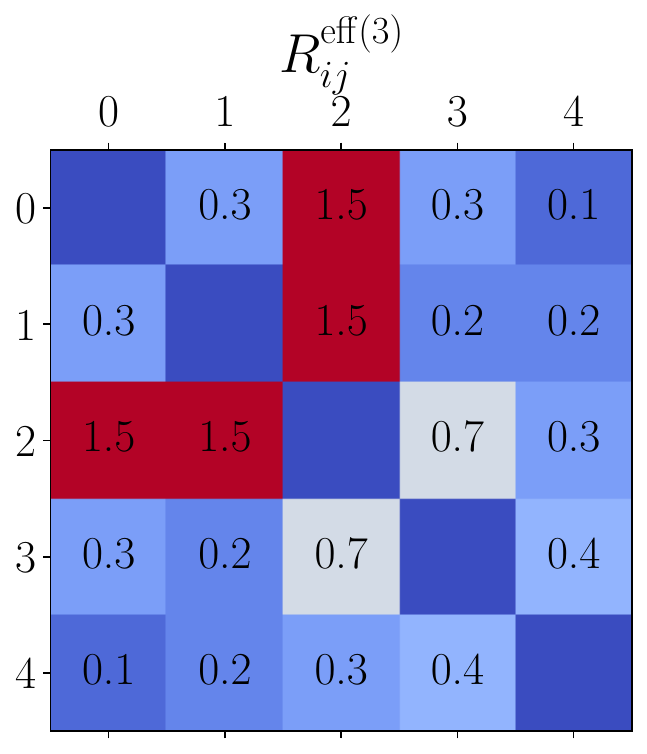}
        \caption{}
        \label{fig:network_5v_iter_3}
    \end{subfigure}
    \begin{subfigure}[t!]{0.45\linewidth}\centering
        \includegraphics[width=0.99\linewidth]{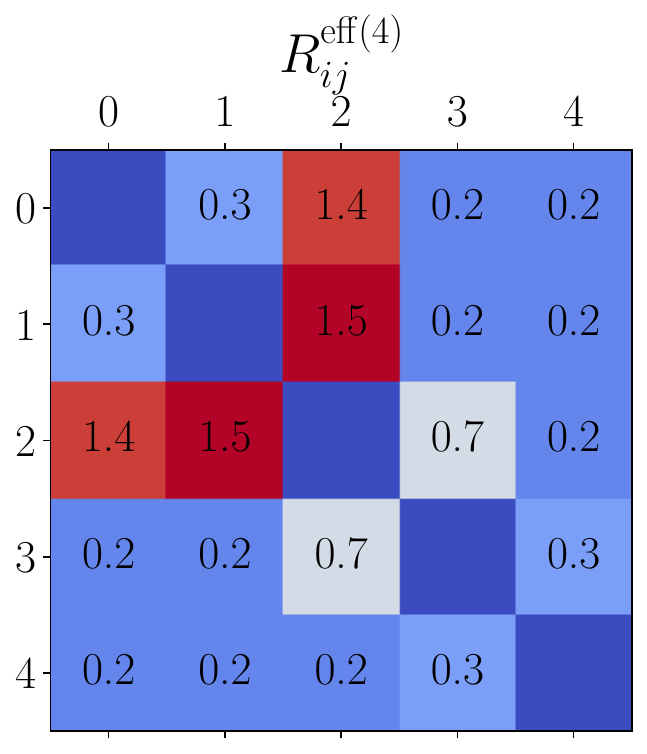}
        \caption{}
        \label{fig:network_5v_iter_4}
    \end{subfigure}
    \caption{(a) The network graph with edges, labeled with secret key generation rates. (b) The initial key rate matrix. (c)--(f) The resulting effective secret key generation rates after every iteration.}
    \label{fig:network_5v}
\end{figure}

To illustrate a step-by-step workflow of the Algorithm~\ref{alg}, we consider a simple network of five nodes, depicted in Fig.~\ref{fig:network_5v_graph}. The graph edges are labeled with corresponding secret key generation rates which are represented by the initial key rate matrix in Fig.~\ref{fig:network_5v_matrix_in}. The target rates are assumed to be $T_{ij}=0.2$. The iteration step is set to be $\delta R=0.1$. 

\begin{enumerate}
    \item At the start the remote nodes apparently have the ``worst'' node pair deficiency,
        \begin{equation*}
            \argmax_{i,j\in V}D(i,j) = \{(0,4),(1,3)\} \,.
        \end{equation*}
        Hence, a random candidate, e.g. $(1,3)$, is chosen. Then the list of all two non-overlapping path combinations from node 1 to node 3 is constructed,
        
        \begin{equation*}
            \begin{alignedat}{2}
                {\cal L}^{2}_{13} = [
                & \{(1, 0, 2, 3), (1, 4, 3)\} \,,  &\quad &\textbf{D}=-0.2 \\
                & \{(1, 0, 3), (1, 4, 2, 3)\} \,,  &\quad &\textbf{D}=-0.1 \\
                & \{(1, 0, 3), (1, 4, 3)\} \,,  &\quad &\textbf{D}=-0.2 \\
                & \{(1, 0, 3), (1, 2, 4, 3)\} \,,  &\quad &\textbf{D}=-0.1 \\
                & \{(1, 0, 3), (1, 2, 3)\} \,,  &\quad &\textbf{D}=-0.3 \\
                & \{(1, 4, 3), (1, 2, 0, 3)\} \,,  &\quad &\textbf{D}=-0.2 \\
                & \{(1, 4, 3), (1, 2, 3)]\} \,. &\quad &\textbf{D}=-0.2
            \end{alignedat}
        \end{equation*}
        Here, in the right column, we put values of ${\bf D}$ for each of combination of paths.
        It is computed as a maximal defficiecy (target rate minus current rate) over all links in the both paths.
        For example, for $\{(1,0,3),(1,4,3)\}$, we have ${\bf D}=\max(0.2-0.5, 0.2-0.5, 0.2-0.4, 0.2-0.6)=-0.2$.
        The lowest value of the path deficiency function, $\textbf{D}=-0.3$, determines the optimal 2--path combination,
        \begin{equation*}
            {\cal P}_{13}^{2,\text{opt}} = \{(1, 0, 3), (1, 2, 3)\} \,.
        \end{equation*}
        The effective rates for all node pairs from ${\cal P}_{13}^{2,\text{opt}}$ are computed as follows:
        \begin{equation*}
            \begin{aligned}
                R_{13}^\eff &= \delta R \,, \\
                R_{ij}^\eff &= R_{ij} - \delta R ~~\forall ij \in\{01,03,12,23\} \,.
            \end{aligned}
        \end{equation*}
        The result can be seen in Fig.~\ref{fig:network_5v_iter_1}.
        The routing list, which was initially empty, is appended with the first entry:
        \begin{equation*}
            {\sf routing\_list} = [\{(1, 0, 3), (1, 2, 3)\}:0.1] \,.
        \end{equation*}

    \item After the first iteration, $(0,4)$ automatically becomes the ``worst'' pair, for which the list of non-overlapping paths with corresponding 2--path deficiencies is
        \begin{equation*}
            \begin{alignedat}{2}
                {\cal L}^{2}_{04} = [
                & \{(0, 1, 4), (0, 2, 4)\} \,, & \quad & \textbf{D}=-0.1 \\
                & \{(0, 1, 4), (0, 2, 3, 4)\} \,, & \quad & \textbf{D}=-0.2 \\
                & \{(0, 1, 4), (0, 3, 4)\} \,, & \quad & \textbf{D}=-0.2 \\
                & \{(0, 1, 4), (0, 3, 2, 4)\} \,, & \quad & \textbf{D}=-0.1 \\
                & \{(0, 1, 2, 4), (0, 3, 4)\} \,, & \quad & \textbf{D}=-0.1 \\
                & \{(0, 2, 4), (0, 3, 4)\} \,, & \quad & \textbf{D}=-0.1 \\
                & \{(0, 2, 1, 4), (0, 3, 4)] \,, & \quad & \textbf{D}=-0.2
            \end{alignedat}
        \end{equation*}
        One can see that there are three candidates with lowest value $\textbf{D}=-0.2$ which form the set ${\cal P}_{04}^{2,\text{pre-opt}}$, but only one has the lowest distance (i.e. the total number of edges that form two paths), equal to 4. Hence, 
        \begin{equation*}
            {\cal P}_{04}^{2,\text{opt}} = \{(0, 1, 4), (0, 3, 4)\} \,.
        \end{equation*}
        The effective rates for all node pairs from ${\cal P}_{04}^{2,\text{opt}}$ are recomputed as follows:
        \begin{equation*}
            \begin{aligned}
                R_{04}^\eff &= \delta R \,, \\
                R_{01(03)}^\eff &\to R_{01(03)}^\eff - \delta R \,, \\
                R_{14(34)}^\eff &= R_{14(34)} - \delta R \,.
            \end{aligned}
        \end{equation*}
        The result can be seen in Fig.~\ref{fig:network_5v_iter_2}.
        The routing list is appended with the second entry:
        \begin{equation*}
            \begin{aligned}
                {\sf routing\_list} = [
                \{(1, 0, 3), (1, 2, 3)\} &: 0.1\,,\\
                \{(0, 1, 4), (0, 3, 4)\} &: 0.1]\,.\\
            \end{aligned}
        \end{equation*}

    \item After the second iteration there are again two candidates with ``worst'' node pair deficiency, $\argmax_{i,j\in V}D(i,j)=\{(0,4),(1,3)\}$, among which, e.g. $(1,3)$, is chosen. The non-overlapping path list is apparently the same as for the first iteration, however, the deficiencies are renewed:
        \begin{equation*}
            \begin{alignedat}{2}
                {\cal L}^{2}_{13} = [
                & \{(1, 0, 2, 3), (1, 4, 3)\} \,, &\quad &\textbf{D}=-0.1 \\
                & \{(1, 0, 3), (1, 4, 2, 3)\} \,, &\quad &\textbf{D}=-0.1 \\
                & \{(1, 0, 3), (1, 4, 3)\} \,, &\quad &\textbf{D}=-0.1 \\
                & \{(1, 0, 3), (1, 2, 4, 3)\} \,, &\quad &\textbf{D}=-0.1 \\
                & \{(1, 0, 3), (1, 2, 3)\} \,, &\quad &\textbf{D}=-0.1 \\
                & \{(1, 4, 3), (1, 2, 0, 3)\} \,, &\quad &\textbf{D}=-0.1 \\
                & \{(1, 4, 3), (1, 2, 3)] \,, &\quad &\textbf{D}=-0.1
            \end{alignedat}
        \end{equation*}
        Among three candidates with lowest distance a random one is chosen:
        \begin{equation*}
            {\cal P}_{13}^{2,\text{opt}} = \{(1, 4, 3), (1, 2, 3)\} \,.
        \end{equation*}
        The new effective rates are
        \begin{equation*}
            \begin{aligned}
                R_{13}^\eff &\to R_{13}^\eff + \delta R \,, \\
                R_{ij}^\eff &\to R_{ij}^\eff - \delta R ~~\forall ij \in \{14,34,12,23\} \,.
            \end{aligned}
        \end{equation*}
        The result can be seen in Fig.~\ref{fig:network_5v_iter_3}.
        The routing list is appended with new entry:
        \begin{equation*}
            \begin{aligned}
                {\sf routing\_list} = [
                \{(1, 0, 3), (1, 2, 3)\} &: 0.1\,,\\
                \{(1, 4, 3), (1, 2, 3)\} &: 0.1\,,\\
                \{(0, 1, 4), (0, 3, 4)\} &: 0.1]\,.\\
            \end{aligned}
        \end{equation*}

    \item After the third iteration $D(0,4)=0.1$ turns out to be the highest. Among all paths from node 0 to node 4,
        \begin{equation*}
            \begin{alignedat}{2}
                {\cal L}^{2}_{04} = [
                & \{(0, 1, 4), (0, 2, 4)\} \,, & \quad & \textbf{D}=0 \\
                & \{(0, 1, 4), (0, 2, 3, 4)\} \,, & \quad & \textbf{D}=0 \\
                & \{(0, 1, 4), (0, 3, 4)\} \,, & \quad & \textbf{D}=0 \\
                & \{(0, 1, 4), (0, 3, 2, 4)\} \,, & \quad & \textbf{D}=0 \\
                & \{(0, 1, 2, 4), (0, 3, 4)\} \,, & \quad & \textbf{D}=-0.1 \\
                & \{(0, 2, 4), (0, 3, 4)\} \,, & \quad & \textbf{D}=-0.1 \\
                & \{(0, 2, 1, 4), (0, 3, 4)] \,, & \quad & \textbf{D}=0
            \end{alignedat}
        \end{equation*}
        there are two candidates with lowest 2--path deficiency $\textbf{D}=-0.1$. The optimal path combination with the lowest distance is
        \begin{equation*}
            {\cal P}_{04}^{2,\text{opt}} = \{(0, 2, 4), (0, 3, 4)\} \,,
        \end{equation*}
        In this way the final rate correction is
        \begin{equation*}
            \begin{aligned}
                R_{04}^\eff &\to R_{04}^\eff + \delta R \,, \\
                R_{ij}^\eff &\to R_{ij}^\eff - \delta R ~~\forall ij \in \{02,24,03,34\} \,.
            \end{aligned}
        \end{equation*}
        The result can be seen in Fig.~\ref{fig:network_5v_iter_3}. 
        The resulting routing list takes the form:
        \begin{equation*}
            \begin{aligned}
                {\sf routing\_list} = [
                \{(1, 0, 3), (1, 2, 3)\} &: 0.1\,,\\
                \{(1, 4, 3), (1, 2, 3)\} &: 0.1\,,\\
                \{(0, 1, 4), (0, 3, 4)\} &: 0.1\,,\\
                \{(0, 2, 4), (0, 3, 4)\} &: 0.1]\,.\\
            \end{aligned}
        \end{equation*}
        Since now all effective rates are not less than the target rate, the algorithm stops.
\end{enumerate}

\end{appendices}

\bibliographystyle{apsrev4-2}
\bibliography{bibliography-qkd.bib}

\begin{thebibliography}{47}%
\makeatletter
\providecommand \@ifxundefined [1]{%
 \@ifx{#1\undefined}
}%
\providecommand \@ifnum [1]{%
 \ifnum #1\expandafter \@firstoftwo
 \else \expandafter \@secondoftwo
 \fi
}%
\providecommand \@ifx [1]{%
 \ifx #1\expandafter \@firstoftwo
 \else \expandafter \@secondoftwo
 \fi
}%
\providecommand \natexlab [1]{#1}%
\providecommand \enquote  [1]{``#1''}%
\providecommand \bibnamefont  [1]{#1}%
\providecommand \bibfnamefont [1]{#1}%
\providecommand \citenamefont [1]{#1}%
\providecommand \href@noop [0]{\@secondoftwo}%
\providecommand \href [0]{\begingroup \@sanitize@url \@href}%
\providecommand \@href[1]{\@@startlink{#1}\@@href}%
\providecommand \@@href[1]{\endgroup#1\@@endlink}%
\providecommand \@sanitize@url [0]{\catcode `\\12\catcode `\$12\catcode `\&12\catcode `\#12\catcode `\^12\catcode `\_12\catcode `\%12\relax}%
\providecommand \@@startlink[1]{}%
\providecommand \@@endlink[0]{}%
\providecommand \url  [0]{\begingroup\@sanitize@url \@url }%
\providecommand \@url [1]{\endgroup\@href {#1}{\urlprefix }}%
\providecommand \urlprefix  [0]{URL }%
\providecommand \Eprint [0]{\href }%
\providecommand \doibase [0]{https://doi.org/}%
\providecommand \selectlanguage [0]{\@gobble}%
\providecommand \bibinfo  [0]{\@secondoftwo}%
\providecommand \bibfield  [0]{\@secondoftwo}%
\providecommand \translation [1]{[#1]}%
\providecommand \BibitemOpen [0]{}%
\providecommand \bibitemStop [0]{}%
\providecommand \bibitemNoStop [0]{.\EOS\space}%
\providecommand \EOS [0]{\spacefactor3000\relax}%
\providecommand \BibitemShut  [1]{\csname bibitem#1\endcsname}%
\let\auto@bib@innerbib\@empty
\bibitem [{\citenamefont {Gisin}\ \emph {et~al.}(2002)\citenamefont {Gisin}, \citenamefont {Ribordy}, \citenamefont {Tittel},\ and\ \citenamefont {Zbinden}}]{Gisin2001}%
  \BibitemOpen
  \bibfield  {author} {\bibinfo {author} {\bibfnamefont {N.}~\bibnamefont {Gisin}}, \bibinfo {author} {\bibfnamefont {G.}~\bibnamefont {Ribordy}}, \bibinfo {author} {\bibfnamefont {W.}~\bibnamefont {Tittel}},\ and\ \bibinfo {author} {\bibfnamefont {H.}~\bibnamefont {Zbinden}},\ }\href {https://doi.org/10.1103/RevModPhys.74.145} {\bibfield  {journal} {\bibinfo  {journal} {Rev. Mod. Phys.}\ }\textbf {\bibinfo {volume} {74}},\ \bibinfo {pages} {145} (\bibinfo {year} {2002})}\BibitemShut {NoStop}%
\bibitem [{\citenamefont {Scarani}\ \emph {et~al.}(2009)\citenamefont {Scarani}, \citenamefont {Bechmann-Pasquinucci}, \citenamefont {Cerf}, \citenamefont {Du\ifmmode~\check{s}\else \v{s}\fi{}ek}, \citenamefont {L\"utkenhaus},\ and\ \citenamefont {Peev}}]{Scarani2009}%
  \BibitemOpen
  \bibfield  {author} {\bibinfo {author} {\bibfnamefont {V.}~\bibnamefont {Scarani}}, \bibinfo {author} {\bibfnamefont {H.}~\bibnamefont {Bechmann-Pasquinucci}}, \bibinfo {author} {\bibfnamefont {N.~J.}\ \bibnamefont {Cerf}}, \bibinfo {author} {\bibfnamefont {M.}~\bibnamefont {Du\ifmmode~\check{s}\else \v{s}\fi{}ek}}, \bibinfo {author} {\bibfnamefont {N.}~\bibnamefont {L\"utkenhaus}},\ and\ \bibinfo {author} {\bibfnamefont {M.}~\bibnamefont {Peev}},\ }\href {https://doi.org/10.1103/RevModPhys.81.1301} {\bibfield  {journal} {\bibinfo  {journal} {Rev. Mod. Phys.}\ }\textbf {\bibinfo {volume} {81}},\ \bibinfo {pages} {1301} (\bibinfo {year} {2009})}\BibitemShut {NoStop}%
\bibitem [{\citenamefont {Lo}\ \emph {et~al.}(2014)\citenamefont {Lo}, \citenamefont {Curty},\ and\ \citenamefont {Tamaki}}]{Lo2014}%
  \BibitemOpen
  \bibfield  {author} {\bibinfo {author} {\bibfnamefont {H.-K.}\ \bibnamefont {Lo}}, \bibinfo {author} {\bibfnamefont {M.}~\bibnamefont {Curty}},\ and\ \bibinfo {author} {\bibfnamefont {K.}~\bibnamefont {Tamaki}},\ }\href {https://doi.org/10.1038/nphoton.2014.149} {\bibfield  {journal} {\bibinfo  {journal} {Nature Photonics}\ }\textbf {\bibinfo {volume} {8}},\ \bibinfo {pages} {595} (\bibinfo {year} {2014})}\BibitemShut {NoStop}%
\bibitem [{\citenamefont {Shor}(1999)}]{Shor1999}%
  \BibitemOpen
  \bibfield  {author} {\bibinfo {author} {\bibfnamefont {P.~W.}\ \bibnamefont {Shor}},\ }\href {https://doi.org/10.1137/S0036144598347011} {\bibfield  {journal} {\bibinfo  {journal} {SIAM Review}\ }\textbf {\bibinfo {volume} {41}},\ \bibinfo {pages} {303} (\bibinfo {year} {1999})}\BibitemShut {NoStop}%
\bibitem [{\citenamefont {Diamanti}\ \emph {et~al.}(2016)\citenamefont {Diamanti}, \citenamefont {Lo}, \citenamefont {Qi},\ and\ \citenamefont {Yuan}}]{Diamanti2016}%
  \BibitemOpen
  \bibfield  {author} {\bibinfo {author} {\bibfnamefont {E.}~\bibnamefont {Diamanti}}, \bibinfo {author} {\bibfnamefont {H.-K.}\ \bibnamefont {Lo}}, \bibinfo {author} {\bibfnamefont {B.}~\bibnamefont {Qi}},\ and\ \bibinfo {author} {\bibfnamefont {Z.}~\bibnamefont {Yuan}},\ }\href {https://doi.org/10.1038/npjqi.2016.25} {\bibfield  {journal} {\bibinfo  {journal} {npj Quantum Information}\ }\textbf {\bibinfo {volume} {2}},\ \bibinfo {pages} {16025} (\bibinfo {year} {2016})}\BibitemShut {NoStop}%
\bibitem [{\citenamefont {Zhang}\ \emph {et~al.}(2018)\citenamefont {Zhang}, \citenamefont {Xu}, \citenamefont {Chen}, \citenamefont {Peng},\ and\ \citenamefont {Pan}}]{Pan2018}%
  \BibitemOpen
  \bibfield  {author} {\bibinfo {author} {\bibfnamefont {Q.}~\bibnamefont {Zhang}}, \bibinfo {author} {\bibfnamefont {F.}~\bibnamefont {Xu}}, \bibinfo {author} {\bibfnamefont {Y.-A.}\ \bibnamefont {Chen}}, \bibinfo {author} {\bibfnamefont {C.-Z.}\ \bibnamefont {Peng}},\ and\ \bibinfo {author} {\bibfnamefont {J.-W.}\ \bibnamefont {Pan}},\ }\href {https://doi.org/10.1364/OE.26.024260} {\bibfield  {journal} {\bibinfo  {journal} {Opt. Express}\ }\textbf {\bibinfo {volume} {26}},\ \bibinfo {pages} {24260} (\bibinfo {year} {2018})}\BibitemShut {NoStop}%
\bibitem [{\citenamefont {Briegel}\ \emph {et~al.}(1998)\citenamefont {Briegel}, \citenamefont {D\"ur}, \citenamefont {Cirac},\ and\ \citenamefont {Zoller}}]{Zoller1998}%
  \BibitemOpen
  \bibfield  {author} {\bibinfo {author} {\bibfnamefont {H.-J.}\ \bibnamefont {Briegel}}, \bibinfo {author} {\bibfnamefont {W.}~\bibnamefont {D\"ur}}, \bibinfo {author} {\bibfnamefont {J.~I.}\ \bibnamefont {Cirac}},\ and\ \bibinfo {author} {\bibfnamefont {P.}~\bibnamefont {Zoller}},\ }\href {https://doi.org/10.1103/PhysRevLett.81.5932} {\bibfield  {journal} {\bibinfo  {journal} {Phys. Rev. Lett.}\ }\textbf {\bibinfo {volume} {81}},\ \bibinfo {pages} {5932} (\bibinfo {year} {1998})}\BibitemShut {NoStop}%
\bibitem [{\citenamefont {Duan}\ \emph {et~al.}(2001)\citenamefont {Duan}, \citenamefont {Lukin}, \citenamefont {Cirac},\ and\ \citenamefont {Zoller}}]{Lukin2001}%
  \BibitemOpen
  \bibfield  {author} {\bibinfo {author} {\bibfnamefont {L.~M.}\ \bibnamefont {Duan}}, \bibinfo {author} {\bibfnamefont {M.~D.}\ \bibnamefont {Lukin}}, \bibinfo {author} {\bibfnamefont {J.~I.}\ \bibnamefont {Cirac}},\ and\ \bibinfo {author} {\bibfnamefont {P.}~\bibnamefont {Zoller}},\ }\href {https://doi.org/10.1038/35106500} {\bibfield  {journal} {\bibinfo  {journal} {Nature}\ }\textbf {\bibinfo {volume} {414}},\ \bibinfo {pages} {413} (\bibinfo {year} {2001})}\BibitemShut {NoStop}%
\bibitem [{\citenamefont {Sangouard}\ \emph {et~al.}(2011)\citenamefont {Sangouard}, \citenamefont {Simon}, \citenamefont {de~Riedmatten},\ and\ \citenamefont {Gisin}}]{Gisin2011}%
  \BibitemOpen
  \bibfield  {author} {\bibinfo {author} {\bibfnamefont {N.}~\bibnamefont {Sangouard}}, \bibinfo {author} {\bibfnamefont {C.}~\bibnamefont {Simon}}, \bibinfo {author} {\bibfnamefont {H.}~\bibnamefont {de~Riedmatten}},\ and\ \bibinfo {author} {\bibfnamefont {N.}~\bibnamefont {Gisin}},\ }\href {https://doi.org/10.1103/RevModPhys.83.33} {\bibfield  {journal} {\bibinfo  {journal} {Rev. Mod. Phys.}\ }\textbf {\bibinfo {volume} {83}},\ \bibinfo {pages} {33} (\bibinfo {year} {2011})}\BibitemShut {NoStop}%
\bibitem [{\citenamefont {Bhaskar}\ \emph {et~al.}(2020)\citenamefont {Bhaskar}, \citenamefont {Riedinger}, \citenamefont {Machielse}, \citenamefont {Levonian}, \citenamefont {Nguyen}, \citenamefont {Knall}, \citenamefont {Park}, \citenamefont {Englund}, \citenamefont {Lon{\v c}ar}, \citenamefont {Sukachev},\ and\ \citenamefont {Lukin}}]{Lukin2020}%
  \BibitemOpen
  \bibfield  {author} {\bibinfo {author} {\bibfnamefont {M.~K.}\ \bibnamefont {Bhaskar}}, \bibinfo {author} {\bibfnamefont {R.}~\bibnamefont {Riedinger}}, \bibinfo {author} {\bibfnamefont {B.}~\bibnamefont {Machielse}}, \bibinfo {author} {\bibfnamefont {D.~S.}\ \bibnamefont {Levonian}}, \bibinfo {author} {\bibfnamefont {C.~T.}\ \bibnamefont {Nguyen}}, \bibinfo {author} {\bibfnamefont {E.~N.}\ \bibnamefont {Knall}}, \bibinfo {author} {\bibfnamefont {H.}~\bibnamefont {Park}}, \bibinfo {author} {\bibfnamefont {D.}~\bibnamefont {Englund}}, \bibinfo {author} {\bibfnamefont {M.}~\bibnamefont {Lon{\v c}ar}}, \bibinfo {author} {\bibfnamefont {D.~D.}\ \bibnamefont {Sukachev}},\ and\ \bibinfo {author} {\bibfnamefont {M.~D.}\ \bibnamefont {Lukin}},\ }\href {https://doi.org/10.1038/s41586-020-2103-5} {\bibfield  {journal} {\bibinfo  {journal} {Nature}\ }\textbf {\bibinfo {volume} {580}},\ \bibinfo {pages} {60} (\bibinfo {year} {2020})}\BibitemShut {NoStop}%
\bibitem [{\citenamefont {Holz\"apfel}\ \emph {et~al.}(2020)\citenamefont {Holz\"apfel}, \citenamefont {Etesse}, \citenamefont {Kaczmarek}, \citenamefont {Tiranov}, \citenamefont {Gisin},\ and\ \citenamefont {Afzelius}}]{Gisin2020}%
  \BibitemOpen
  \bibfield  {author} {\bibinfo {author} {\bibfnamefont {A.}~\bibnamefont {Holz\"apfel}}, \bibinfo {author} {\bibfnamefont {J.}~\bibnamefont {Etesse}}, \bibinfo {author} {\bibfnamefont {K.~T.}\ \bibnamefont {Kaczmarek}}, \bibinfo {author} {\bibfnamefont {A.}~\bibnamefont {Tiranov}}, \bibinfo {author} {\bibfnamefont {N.}~\bibnamefont {Gisin}},\ and\ \bibinfo {author} {\bibfnamefont {M.}~\bibnamefont {Afzelius}},\ }\href {https://doi.org/10.1088/1367-2630/ab8aac} {\bibfield  {journal} {\bibinfo  {journal} {New Journal of Physics}\ }\textbf {\bibinfo {volume} {22}},\ \bibinfo {pages} {063009} (\bibinfo {year} {2020})}\BibitemShut {NoStop}%
\bibitem [{\citenamefont {Bersin}\ \emph {et~al.}(2024)\citenamefont {Bersin}, \citenamefont {Sutula}, \citenamefont {Huan}, \citenamefont {Suleymanzade}, \citenamefont {Assumpcao}, \citenamefont {Wei}, \citenamefont {Stas}, \citenamefont {Knaut}, \citenamefont {Knall}, \citenamefont {Langrock}, \citenamefont {Sinclair}, \citenamefont {Murphy}, \citenamefont {Riedinger}, \citenamefont {Yeh}, \citenamefont {Xin}, \citenamefont {Bandyopadhyay}, \citenamefont {Sukachev}, \citenamefont {Machielse}, \citenamefont {Levonian}, \citenamefont {Bhaskar}, \citenamefont {Hamilton}, \citenamefont {Park}, \citenamefont {Lon\ifmmode~\check{c}\else \v{c}\fi{}ar}, \citenamefont {Fejer}, \citenamefont {Dixon}, \citenamefont {Englund},\ and\ \citenamefont {Lukin}}]{Lukin2024}%
  \BibitemOpen
  \bibfield  {author} {\bibinfo {author} {\bibfnamefont {E.}~\bibnamefont {Bersin}}, \bibinfo {author} {\bibfnamefont {M.}~\bibnamefont {Sutula}}, \bibinfo {author} {\bibfnamefont {Y.~Q.}\ \bibnamefont {Huan}}, \bibinfo {author} {\bibfnamefont {A.}~\bibnamefont {Suleymanzade}}, \bibinfo {author} {\bibfnamefont {D.~R.}\ \bibnamefont {Assumpcao}}, \bibinfo {author} {\bibfnamefont {Y.-C.}\ \bibnamefont {Wei}}, \bibinfo {author} {\bibfnamefont {P.-J.}\ \bibnamefont {Stas}}, \bibinfo {author} {\bibfnamefont {C.~M.}\ \bibnamefont {Knaut}}, \bibinfo {author} {\bibfnamefont {E.~N.}\ \bibnamefont {Knall}}, \bibinfo {author} {\bibfnamefont {C.}~\bibnamefont {Langrock}}, \bibinfo {author} {\bibfnamefont {N.}~\bibnamefont {Sinclair}}, \bibinfo {author} {\bibfnamefont {R.}~\bibnamefont {Murphy}}, \bibinfo {author} {\bibfnamefont {R.}~\bibnamefont {Riedinger}}, \bibinfo {author} {\bibfnamefont {M.}~\bibnamefont {Yeh}}, \bibinfo {author} {\bibfnamefont {C.}~\bibnamefont {Xin}}, \bibinfo {author} {\bibfnamefont
  {S.}~\bibnamefont {Bandyopadhyay}}, \bibinfo {author} {\bibfnamefont {D.~D.}\ \bibnamefont {Sukachev}}, \bibinfo {author} {\bibfnamefont {B.}~\bibnamefont {Machielse}}, \bibinfo {author} {\bibfnamefont {D.~S.}\ \bibnamefont {Levonian}}, \bibinfo {author} {\bibfnamefont {M.~K.}\ \bibnamefont {Bhaskar}}, \bibinfo {author} {\bibfnamefont {S.}~\bibnamefont {Hamilton}}, \bibinfo {author} {\bibfnamefont {H.}~\bibnamefont {Park}}, \bibinfo {author} {\bibfnamefont {M.}~\bibnamefont {Lon\ifmmode~\check{c}\else \v{c}\fi{}ar}}, \bibinfo {author} {\bibfnamefont {M.~M.}\ \bibnamefont {Fejer}}, \bibinfo {author} {\bibfnamefont {P.~B.}\ \bibnamefont {Dixon}}, \bibinfo {author} {\bibfnamefont {D.~R.}\ \bibnamefont {Englund}},\ and\ \bibinfo {author} {\bibfnamefont {M.~D.}\ \bibnamefont {Lukin}},\ }\href {https://doi.org/10.1103/PRXQuantum.5.010303} {\bibfield  {journal} {\bibinfo  {journal} {PRX Quantum}\ }\textbf {\bibinfo {volume} {5}},\ \bibinfo {pages} {010303} (\bibinfo {year} {2024})}\BibitemShut {NoStop}%
\bibitem [{\citenamefont {Knaut}\ \emph {et~al.}(2024)\citenamefont {Knaut}, \citenamefont {Suleymanzade}, \citenamefont {Wei}, \citenamefont {Assumpcao}, \citenamefont {Stas}, \citenamefont {Huan}, \citenamefont {Machielse}, \citenamefont {Knall}, \citenamefont {Sutula}, \citenamefont {Baranes}, \citenamefont {Sinclair}, \citenamefont {De-Eknamkul}, \citenamefont {Levonian}, \citenamefont {Bhaskar}, \citenamefont {Park}, \citenamefont {Lon{\v c}ar},\ and\ \citenamefont {Lukin}}]{Lukin20242}%
  \BibitemOpen
  \bibfield  {author} {\bibinfo {author} {\bibfnamefont {C.~M.}\ \bibnamefont {Knaut}}, \bibinfo {author} {\bibfnamefont {A.}~\bibnamefont {Suleymanzade}}, \bibinfo {author} {\bibfnamefont {Y.~C.}\ \bibnamefont {Wei}}, \bibinfo {author} {\bibfnamefont {D.~R.}\ \bibnamefont {Assumpcao}}, \bibinfo {author} {\bibfnamefont {P.~J.}\ \bibnamefont {Stas}}, \bibinfo {author} {\bibfnamefont {Y.~Q.}\ \bibnamefont {Huan}}, \bibinfo {author} {\bibfnamefont {B.}~\bibnamefont {Machielse}}, \bibinfo {author} {\bibfnamefont {E.~N.}\ \bibnamefont {Knall}}, \bibinfo {author} {\bibfnamefont {M.}~\bibnamefont {Sutula}}, \bibinfo {author} {\bibfnamefont {G.}~\bibnamefont {Baranes}}, \bibinfo {author} {\bibfnamefont {N.}~\bibnamefont {Sinclair}}, \bibinfo {author} {\bibfnamefont {C.}~\bibnamefont {De-Eknamkul}}, \bibinfo {author} {\bibfnamefont {D.~S.}\ \bibnamefont {Levonian}}, \bibinfo {author} {\bibfnamefont {M.~K.}\ \bibnamefont {Bhaskar}}, \bibinfo {author} {\bibfnamefont {H.}~\bibnamefont {Park}}, \bibinfo {author}
  {\bibfnamefont {M.}~\bibnamefont {Lon{\v c}ar}},\ and\ \bibinfo {author} {\bibfnamefont {M.~D.}\ \bibnamefont {Lukin}},\ }\href {https://doi.org/10.1038/s41586-024-07252-z} {\bibfield  {journal} {\bibinfo  {journal} {Nature}\ }\textbf {\bibinfo {volume} {629}},\ \bibinfo {pages} {573} (\bibinfo {year} {2024})}\BibitemShut {NoStop}%
\bibitem [{\citenamefont {Elliott}(2004)}]{Elliott2004}%
  \BibitemOpen
  \bibfield  {author} {\bibinfo {author} {\bibfnamefont {C.}~\bibnamefont {Elliott}},\ }\href {https://arxiv.org/abs/quant-ph/0412029} {\bibinfo {title} {The darpa quantum network}} (\bibinfo {year} {2004}),\ \Eprint {https://arxiv.org/abs/quant-ph/0412029} {arXiv:quant-ph/0412029 [quant-ph]} \BibitemShut {NoStop}%
\bibitem [{\citenamefont {Peev}\ \emph {et~al.}(2009)\citenamefont {Peev}, \citenamefont {Pacher}, \citenamefont {All{\'e}aume}, \citenamefont {Barreiro}, \citenamefont {Bouda}, \citenamefont {Boxleitner}, \citenamefont {Debuisschert}, \citenamefont {Diamanti}, \citenamefont {Dianati}, \citenamefont {Dynes}, \citenamefont {Fasel}, \citenamefont {Fossier}, \citenamefont {F{\"u}rst}, \citenamefont {Gautier}, \citenamefont {Gay}, \citenamefont {Gisin}, \citenamefont {Grangier}, \citenamefont {Happe}, \citenamefont {Hasani}, \citenamefont {Hentschel}, \citenamefont {H{\"u}bel}, \citenamefont {Humer}, \citenamefont {L{\"a}nger}, \citenamefont {Legr{\'e}}, \citenamefont {Lieger}, \citenamefont {Lodewyck}, \citenamefont {Lor{\"u}nser}, \citenamefont {L{\"u}tkenhaus}, \citenamefont {Marhold}, \citenamefont {Matyus}, \citenamefont {Maurhart}, \citenamefont {Monat}, \citenamefont {Nauerth}, \citenamefont {Page}, \citenamefont {Poppe}, \citenamefont {Querasser}, \citenamefont {Ribordy}, \citenamefont {Robyr},
  \citenamefont {Salvail}, \citenamefont {Sharpe}, \citenamefont {Shields}, \citenamefont {Stucki}, \citenamefont {Suda}, \citenamefont {Tamas}, \citenamefont {Themel}, \citenamefont {Thew}, \citenamefont {Thoma}, \citenamefont {Treiber}, \citenamefont {Trinkler}, \citenamefont {Tualle-Brouri}, \citenamefont {Vannel}, \citenamefont {Walenta}, \citenamefont {Weier}, \citenamefont {Weinfurter}, \citenamefont {Wimberger}, \citenamefont {Yuan}, \citenamefont {Zbinden},\ and\ \citenamefont {Zeilinger}}]{Peev2009}%
  \BibitemOpen
  \bibfield  {author} {\bibinfo {author} {\bibfnamefont {M.}~\bibnamefont {Peev}}, \bibinfo {author} {\bibfnamefont {C.}~\bibnamefont {Pacher}}, \bibinfo {author} {\bibfnamefont {R.}~\bibnamefont {All{\'e}aume}}, \bibinfo {author} {\bibfnamefont {C.}~\bibnamefont {Barreiro}}, \bibinfo {author} {\bibfnamefont {J.}~\bibnamefont {Bouda}}, \bibinfo {author} {\bibfnamefont {W.}~\bibnamefont {Boxleitner}}, \bibinfo {author} {\bibfnamefont {T.}~\bibnamefont {Debuisschert}}, \bibinfo {author} {\bibfnamefont {E.}~\bibnamefont {Diamanti}}, \bibinfo {author} {\bibfnamefont {M.}~\bibnamefont {Dianati}}, \bibinfo {author} {\bibfnamefont {J.~F.}\ \bibnamefont {Dynes}}, \bibinfo {author} {\bibfnamefont {S.}~\bibnamefont {Fasel}}, \bibinfo {author} {\bibfnamefont {S.}~\bibnamefont {Fossier}}, \bibinfo {author} {\bibfnamefont {M.}~\bibnamefont {F{\"u}rst}}, \bibinfo {author} {\bibfnamefont {J.-D.}\ \bibnamefont {Gautier}}, \bibinfo {author} {\bibfnamefont {O.}~\bibnamefont {Gay}}, \bibinfo {author} {\bibfnamefont
  {N.}~\bibnamefont {Gisin}}, \bibinfo {author} {\bibfnamefont {P.}~\bibnamefont {Grangier}}, \bibinfo {author} {\bibfnamefont {A.}~\bibnamefont {Happe}}, \bibinfo {author} {\bibfnamefont {Y.}~\bibnamefont {Hasani}}, \bibinfo {author} {\bibfnamefont {M.}~\bibnamefont {Hentschel}}, \bibinfo {author} {\bibfnamefont {H.}~\bibnamefont {H{\"u}bel}}, \bibinfo {author} {\bibfnamefont {G.}~\bibnamefont {Humer}}, \bibinfo {author} {\bibfnamefont {T.}~\bibnamefont {L{\"a}nger}}, \bibinfo {author} {\bibfnamefont {M.}~\bibnamefont {Legr{\'e}}}, \bibinfo {author} {\bibfnamefont {R.}~\bibnamefont {Lieger}}, \bibinfo {author} {\bibfnamefont {J.}~\bibnamefont {Lodewyck}}, \bibinfo {author} {\bibfnamefont {T.}~\bibnamefont {Lor{\"u}nser}}, \bibinfo {author} {\bibfnamefont {N.}~\bibnamefont {L{\"u}tkenhaus}}, \bibinfo {author} {\bibfnamefont {A.}~\bibnamefont {Marhold}}, \bibinfo {author} {\bibfnamefont {T.}~\bibnamefont {Matyus}}, \bibinfo {author} {\bibfnamefont {O.}~\bibnamefont {Maurhart}}, \bibinfo {author} {\bibfnamefont
  {L.}~\bibnamefont {Monat}}, \bibinfo {author} {\bibfnamefont {S.}~\bibnamefont {Nauerth}}, \bibinfo {author} {\bibfnamefont {J.-B.}\ \bibnamefont {Page}}, \bibinfo {author} {\bibfnamefont {A.}~\bibnamefont {Poppe}}, \bibinfo {author} {\bibfnamefont {E.}~\bibnamefont {Querasser}}, \bibinfo {author} {\bibfnamefont {G.}~\bibnamefont {Ribordy}}, \bibinfo {author} {\bibfnamefont {S.}~\bibnamefont {Robyr}}, \bibinfo {author} {\bibfnamefont {L.}~\bibnamefont {Salvail}}, \bibinfo {author} {\bibfnamefont {A.~W.}\ \bibnamefont {Sharpe}}, \bibinfo {author} {\bibfnamefont {A.~J.}\ \bibnamefont {Shields}}, \bibinfo {author} {\bibfnamefont {D.}~\bibnamefont {Stucki}}, \bibinfo {author} {\bibfnamefont {M.}~\bibnamefont {Suda}}, \bibinfo {author} {\bibfnamefont {C.}~\bibnamefont {Tamas}}, \bibinfo {author} {\bibfnamefont {T.}~\bibnamefont {Themel}}, \bibinfo {author} {\bibfnamefont {R.~T.}\ \bibnamefont {Thew}}, \bibinfo {author} {\bibfnamefont {Y.}~\bibnamefont {Thoma}}, \bibinfo {author} {\bibfnamefont {A.}~\bibnamefont
  {Treiber}}, \bibinfo {author} {\bibfnamefont {P.}~\bibnamefont {Trinkler}}, \bibinfo {author} {\bibfnamefont {R.}~\bibnamefont {Tualle-Brouri}}, \bibinfo {author} {\bibfnamefont {F.}~\bibnamefont {Vannel}}, \bibinfo {author} {\bibfnamefont {N.}~\bibnamefont {Walenta}}, \bibinfo {author} {\bibfnamefont {H.}~\bibnamefont {Weier}}, \bibinfo {author} {\bibfnamefont {H.}~\bibnamefont {Weinfurter}}, \bibinfo {author} {\bibfnamefont {I.}~\bibnamefont {Wimberger}}, \bibinfo {author} {\bibfnamefont {Z.~L.}\ \bibnamefont {Yuan}}, \bibinfo {author} {\bibfnamefont {H.}~\bibnamefont {Zbinden}},\ and\ \bibinfo {author} {\bibfnamefont {A.}~\bibnamefont {Zeilinger}},\ }\href {https://doi.org/10.1088/1367-2630/11/7/075001} {\bibfield  {journal} {\bibinfo  {journal} {New Journal of Physics}\ }\textbf {\bibinfo {volume} {11}},\ \bibinfo {pages} {075001} (\bibinfo {year} {2009})}\BibitemShut {NoStop}%
\bibitem [{\citenamefont {Stucki}\ \emph {et~al.}(2011)\citenamefont {Stucki}, \citenamefont {Legr{\'e}}, \citenamefont {Buntschu}, \citenamefont {Clausen}, \citenamefont {Felber}, \citenamefont {Gisin}, \citenamefont {Henzen}, \citenamefont {Junod}, \citenamefont {Litzistorf}, \citenamefont {Monbaron}, \citenamefont {Monat}, \citenamefont {Page}, \citenamefont {Perroud}, \citenamefont {Ribordy}, \citenamefont {Rochas}, \citenamefont {Robyr}, \citenamefont {Tavares}, \citenamefont {Thew}, \citenamefont {Trinkler}, \citenamefont {Ventura}, \citenamefont {Voirol}, \citenamefont {Walenta},\ and\ \citenamefont {Zbinden}}]{Stucki2011}%
  \BibitemOpen
  \bibfield  {author} {\bibinfo {author} {\bibfnamefont {D.}~\bibnamefont {Stucki}}, \bibinfo {author} {\bibfnamefont {M.}~\bibnamefont {Legr{\'e}}}, \bibinfo {author} {\bibfnamefont {F.}~\bibnamefont {Buntschu}}, \bibinfo {author} {\bibfnamefont {B.}~\bibnamefont {Clausen}}, \bibinfo {author} {\bibfnamefont {N.}~\bibnamefont {Felber}}, \bibinfo {author} {\bibfnamefont {N.}~\bibnamefont {Gisin}}, \bibinfo {author} {\bibfnamefont {L.}~\bibnamefont {Henzen}}, \bibinfo {author} {\bibfnamefont {P.}~\bibnamefont {Junod}}, \bibinfo {author} {\bibfnamefont {G.}~\bibnamefont {Litzistorf}}, \bibinfo {author} {\bibfnamefont {P.}~\bibnamefont {Monbaron}}, \bibinfo {author} {\bibfnamefont {L.}~\bibnamefont {Monat}}, \bibinfo {author} {\bibfnamefont {J.-B.}\ \bibnamefont {Page}}, \bibinfo {author} {\bibfnamefont {D.}~\bibnamefont {Perroud}}, \bibinfo {author} {\bibfnamefont {G.}~\bibnamefont {Ribordy}}, \bibinfo {author} {\bibfnamefont {A.}~\bibnamefont {Rochas}}, \bibinfo {author} {\bibfnamefont {S.}~\bibnamefont
  {Robyr}}, \bibinfo {author} {\bibfnamefont {J.}~\bibnamefont {Tavares}}, \bibinfo {author} {\bibfnamefont {R.}~\bibnamefont {Thew}}, \bibinfo {author} {\bibfnamefont {P.}~\bibnamefont {Trinkler}}, \bibinfo {author} {\bibfnamefont {S.}~\bibnamefont {Ventura}}, \bibinfo {author} {\bibfnamefont {R.}~\bibnamefont {Voirol}}, \bibinfo {author} {\bibfnamefont {N.}~\bibnamefont {Walenta}},\ and\ \bibinfo {author} {\bibfnamefont {H.}~\bibnamefont {Zbinden}},\ }\href {https://doi.org/10.1088/1367-2630/13/12/123001} {\bibfield  {journal} {\bibinfo  {journal} {New Journal of Physics}\ }\textbf {\bibinfo {volume} {13}},\ \bibinfo {pages} {123001} (\bibinfo {year} {2011})}\BibitemShut {NoStop}%
\bibitem [{\citenamefont {Chen}\ \emph {et~al.}(2009)\citenamefont {Chen}, \citenamefont {Liang}, \citenamefont {Liu}, \citenamefont {Cai}, \citenamefont {Ju}, \citenamefont {Liu}, \citenamefont {Wang}, \citenamefont {Yin}, \citenamefont {Chen}, \citenamefont {Chen}, \citenamefont {Peng},\ and\ \citenamefont {Pan}}]{Pan2009}%
  \BibitemOpen
  \bibfield  {author} {\bibinfo {author} {\bibfnamefont {T.-Y.}\ \bibnamefont {Chen}}, \bibinfo {author} {\bibfnamefont {H.}~\bibnamefont {Liang}}, \bibinfo {author} {\bibfnamefont {Y.}~\bibnamefont {Liu}}, \bibinfo {author} {\bibfnamefont {W.-Q.}\ \bibnamefont {Cai}}, \bibinfo {author} {\bibfnamefont {L.}~\bibnamefont {Ju}}, \bibinfo {author} {\bibfnamefont {W.-Y.}\ \bibnamefont {Liu}}, \bibinfo {author} {\bibfnamefont {J.}~\bibnamefont {Wang}}, \bibinfo {author} {\bibfnamefont {H.}~\bibnamefont {Yin}}, \bibinfo {author} {\bibfnamefont {K.}~\bibnamefont {Chen}}, \bibinfo {author} {\bibfnamefont {Z.-B.}\ \bibnamefont {Chen}}, \bibinfo {author} {\bibfnamefont {C.-Z.}\ \bibnamefont {Peng}},\ and\ \bibinfo {author} {\bibfnamefont {J.-W.}\ \bibnamefont {Pan}},\ }\href {https://doi.org/10.1364/OE.17.006540} {\bibfield  {journal} {\bibinfo  {journal} {Opt. Express}\ }\textbf {\bibinfo {volume} {17}},\ \bibinfo {pages} {6540} (\bibinfo {year} {2009})}\BibitemShut {NoStop}%
\bibitem [{\citenamefont {Chen}\ \emph {et~al.}(2010)\citenamefont {Chen}, \citenamefont {Wang}, \citenamefont {Liang}, \citenamefont {Liu}, \citenamefont {Liu}, \citenamefont {Jiang}, \citenamefont {Wang}, \citenamefont {Wan}, \citenamefont {Cai}, \citenamefont {Ju}, \citenamefont {Chen}, \citenamefont {Wang}, \citenamefont {Gao}, \citenamefont {Chen}, \citenamefont {Peng}, \citenamefont {Chen},\ and\ \citenamefont {Pan}}]{Pan2010}%
  \BibitemOpen
  \bibfield  {author} {\bibinfo {author} {\bibfnamefont {T.-Y.}\ \bibnamefont {Chen}}, \bibinfo {author} {\bibfnamefont {J.}~\bibnamefont {Wang}}, \bibinfo {author} {\bibfnamefont {H.}~\bibnamefont {Liang}}, \bibinfo {author} {\bibfnamefont {W.-Y.}\ \bibnamefont {Liu}}, \bibinfo {author} {\bibfnamefont {Y.}~\bibnamefont {Liu}}, \bibinfo {author} {\bibfnamefont {X.}~\bibnamefont {Jiang}}, \bibinfo {author} {\bibfnamefont {Y.}~\bibnamefont {Wang}}, \bibinfo {author} {\bibfnamefont {X.}~\bibnamefont {Wan}}, \bibinfo {author} {\bibfnamefont {W.-Q.}\ \bibnamefont {Cai}}, \bibinfo {author} {\bibfnamefont {L.}~\bibnamefont {Ju}}, \bibinfo {author} {\bibfnamefont {L.-K.}\ \bibnamefont {Chen}}, \bibinfo {author} {\bibfnamefont {L.-J.}\ \bibnamefont {Wang}}, \bibinfo {author} {\bibfnamefont {Y.}~\bibnamefont {Gao}}, \bibinfo {author} {\bibfnamefont {K.}~\bibnamefont {Chen}}, \bibinfo {author} {\bibfnamefont {C.-Z.}\ \bibnamefont {Peng}}, \bibinfo {author} {\bibfnamefont {Z.-B.}\ \bibnamefont {Chen}},\ and\ \bibinfo
  {author} {\bibfnamefont {J.-W.}\ \bibnamefont {Pan}},\ }\href {https://doi.org/10.1364/OE.18.027217} {\bibfield  {journal} {\bibinfo  {journal} {Opt. Express}\ }\textbf {\bibinfo {volume} {18}},\ \bibinfo {pages} {27217} (\bibinfo {year} {2010})}\BibitemShut {NoStop}%
\bibitem [{\citenamefont {Wang}\ \emph {et~al.}(2010)\citenamefont {Wang}, \citenamefont {Chen}, \citenamefont {Yin}, \citenamefont {Zhang}, \citenamefont {Zhang}, \citenamefont {Li}, \citenamefont {Xu}, \citenamefont {Zhou}, \citenamefont {Yang}, \citenamefont {Huang}, \citenamefont {Zhang}, \citenamefont {Li}, \citenamefont {Liu}, \citenamefont {Wang}, \citenamefont {Guo},\ and\ \citenamefont {Han}}]{Han2010}%
  \BibitemOpen
  \bibfield  {author} {\bibinfo {author} {\bibfnamefont {S.}~\bibnamefont {Wang}}, \bibinfo {author} {\bibfnamefont {W.}~\bibnamefont {Chen}}, \bibinfo {author} {\bibfnamefont {Z.-Q.}\ \bibnamefont {Yin}}, \bibinfo {author} {\bibfnamefont {Y.}~\bibnamefont {Zhang}}, \bibinfo {author} {\bibfnamefont {T.}~\bibnamefont {Zhang}}, \bibinfo {author} {\bibfnamefont {H.-W.}\ \bibnamefont {Li}}, \bibinfo {author} {\bibfnamefont {F.-X.}\ \bibnamefont {Xu}}, \bibinfo {author} {\bibfnamefont {Z.}~\bibnamefont {Zhou}}, \bibinfo {author} {\bibfnamefont {Y.}~\bibnamefont {Yang}}, \bibinfo {author} {\bibfnamefont {D.-J.}\ \bibnamefont {Huang}}, \bibinfo {author} {\bibfnamefont {L.-J.}\ \bibnamefont {Zhang}}, \bibinfo {author} {\bibfnamefont {F.-Y.}\ \bibnamefont {Li}}, \bibinfo {author} {\bibfnamefont {D.}~\bibnamefont {Liu}}, \bibinfo {author} {\bibfnamefont {Y.-G.}\ \bibnamefont {Wang}}, \bibinfo {author} {\bibfnamefont {G.-C.}\ \bibnamefont {Guo}},\ and\ \bibinfo {author} {\bibfnamefont {Z.-F.}\ \bibnamefont {Han}},\
  }\href {https://doi.org/10.1364/OL.35.002454} {\bibfield  {journal} {\bibinfo  {journal} {Opt. Lett.}\ }\textbf {\bibinfo {volume} {35}},\ \bibinfo {pages} {2454} (\bibinfo {year} {2010})}\BibitemShut {NoStop}%
\bibitem [{\citenamefont {Sasaki}\ \emph {et~al.}(2011)\citenamefont {Sasaki}, \citenamefont {Fujiwara}, \citenamefont {Ishizuka}, \citenamefont {Klaus}, \citenamefont {Wakui}, \citenamefont {Takeoka}, \citenamefont {Miki}, \citenamefont {Yamashita}, \citenamefont {Wang}, \citenamefont {Tanaka}, \citenamefont {Yoshino}, \citenamefont {Nambu}, \citenamefont {Takahashi}, \citenamefont {Tajima}, \citenamefont {Tomita}, \citenamefont {Domeki}, \citenamefont {Hasegawa}, \citenamefont {Sakai}, \citenamefont {Kobayashi}, \citenamefont {Asai}, \citenamefont {Shimizu}, \citenamefont {Tokura}, \citenamefont {Tsurumaru}, \citenamefont {Matsui}, \citenamefont {Honjo}, \citenamefont {Tamaki}, \citenamefont {Takesue}, \citenamefont {Tokura}, \citenamefont {Dynes}, \citenamefont {Dixon}, \citenamefont {Sharpe}, \citenamefont {Yuan}, \citenamefont {Shields}, \citenamefont {Uchikoga}, \citenamefont {Legr\'{e}}, \citenamefont {Robyr}, \citenamefont {Trinkler}, \citenamefont {Monat}, \citenamefont {Page}, \citenamefont
  {Ribordy}, \citenamefont {Poppe}, \citenamefont {Allacher}, \citenamefont {Maurhart}, \citenamefont {L\"{a}nger}, \citenamefont {Peev},\ and\ \citenamefont {Zeilinger}}]{Zeilinger2011}%
  \BibitemOpen
  \bibfield  {author} {\bibinfo {author} {\bibfnamefont {M.}~\bibnamefont {Sasaki}}, \bibinfo {author} {\bibfnamefont {M.}~\bibnamefont {Fujiwara}}, \bibinfo {author} {\bibfnamefont {H.}~\bibnamefont {Ishizuka}}, \bibinfo {author} {\bibfnamefont {W.}~\bibnamefont {Klaus}}, \bibinfo {author} {\bibfnamefont {K.}~\bibnamefont {Wakui}}, \bibinfo {author} {\bibfnamefont {M.}~\bibnamefont {Takeoka}}, \bibinfo {author} {\bibfnamefont {S.}~\bibnamefont {Miki}}, \bibinfo {author} {\bibfnamefont {T.}~\bibnamefont {Yamashita}}, \bibinfo {author} {\bibfnamefont {Z.}~\bibnamefont {Wang}}, \bibinfo {author} {\bibfnamefont {A.}~\bibnamefont {Tanaka}}, \bibinfo {author} {\bibfnamefont {K.}~\bibnamefont {Yoshino}}, \bibinfo {author} {\bibfnamefont {Y.}~\bibnamefont {Nambu}}, \bibinfo {author} {\bibfnamefont {S.}~\bibnamefont {Takahashi}}, \bibinfo {author} {\bibfnamefont {A.}~\bibnamefont {Tajima}}, \bibinfo {author} {\bibfnamefont {A.}~\bibnamefont {Tomita}}, \bibinfo {author} {\bibfnamefont {T.}~\bibnamefont {Domeki}},
  \bibinfo {author} {\bibfnamefont {T.}~\bibnamefont {Hasegawa}}, \bibinfo {author} {\bibfnamefont {Y.}~\bibnamefont {Sakai}}, \bibinfo {author} {\bibfnamefont {H.}~\bibnamefont {Kobayashi}}, \bibinfo {author} {\bibfnamefont {T.}~\bibnamefont {Asai}}, \bibinfo {author} {\bibfnamefont {K.}~\bibnamefont {Shimizu}}, \bibinfo {author} {\bibfnamefont {T.}~\bibnamefont {Tokura}}, \bibinfo {author} {\bibfnamefont {T.}~\bibnamefont {Tsurumaru}}, \bibinfo {author} {\bibfnamefont {M.}~\bibnamefont {Matsui}}, \bibinfo {author} {\bibfnamefont {T.}~\bibnamefont {Honjo}}, \bibinfo {author} {\bibfnamefont {K.}~\bibnamefont {Tamaki}}, \bibinfo {author} {\bibfnamefont {H.}~\bibnamefont {Takesue}}, \bibinfo {author} {\bibfnamefont {Y.}~\bibnamefont {Tokura}}, \bibinfo {author} {\bibfnamefont {J.~F.}\ \bibnamefont {Dynes}}, \bibinfo {author} {\bibfnamefont {A.~R.}\ \bibnamefont {Dixon}}, \bibinfo {author} {\bibfnamefont {A.~W.}\ \bibnamefont {Sharpe}}, \bibinfo {author} {\bibfnamefont {Z.~L.}\ \bibnamefont {Yuan}}, \bibinfo
  {author} {\bibfnamefont {A.~J.}\ \bibnamefont {Shields}}, \bibinfo {author} {\bibfnamefont {S.}~\bibnamefont {Uchikoga}}, \bibinfo {author} {\bibfnamefont {M.}~\bibnamefont {Legr\'{e}}}, \bibinfo {author} {\bibfnamefont {S.}~\bibnamefont {Robyr}}, \bibinfo {author} {\bibfnamefont {P.}~\bibnamefont {Trinkler}}, \bibinfo {author} {\bibfnamefont {L.}~\bibnamefont {Monat}}, \bibinfo {author} {\bibfnamefont {J.-B.}\ \bibnamefont {Page}}, \bibinfo {author} {\bibfnamefont {G.}~\bibnamefont {Ribordy}}, \bibinfo {author} {\bibfnamefont {A.}~\bibnamefont {Poppe}}, \bibinfo {author} {\bibfnamefont {A.}~\bibnamefont {Allacher}}, \bibinfo {author} {\bibfnamefont {O.}~\bibnamefont {Maurhart}}, \bibinfo {author} {\bibfnamefont {T.}~\bibnamefont {L\"{a}nger}}, \bibinfo {author} {\bibfnamefont {M.}~\bibnamefont {Peev}},\ and\ \bibinfo {author} {\bibfnamefont {A.}~\bibnamefont {Zeilinger}},\ }\href {https://doi.org/10.1364/OE.19.010387} {\bibfield  {journal} {\bibinfo  {journal} {Opt. Express}\ }\textbf {\bibinfo {volume}
  {19}},\ \bibinfo {pages} {10387} (\bibinfo {year} {2011})}\BibitemShut {NoStop}%
\bibitem [{\citenamefont {Alleaume}\ \emph {et~al.}(2009)\citenamefont {Alleaume}, \citenamefont {Roueff}, \citenamefont {Diamanti},\ and\ \citenamefont {L\"utkenhaus}}]{Lutkenhaus2009}%
  \BibitemOpen
  \bibfield  {author} {\bibinfo {author} {\bibfnamefont {R.}~\bibnamefont {Alleaume}}, \bibinfo {author} {\bibfnamefont {F.}~\bibnamefont {Roueff}}, \bibinfo {author} {\bibfnamefont {E.}~\bibnamefont {Diamanti}},\ and\ \bibinfo {author} {\bibfnamefont {N.}~\bibnamefont {L\"utkenhaus}},\ }\href {https://doi.org/10.1088/1367-2630/11/7/075002} {\bibfield  {journal} {\bibinfo  {journal} {New Journal of Physics}\ }\textbf {\bibinfo {volume} {11}},\ \bibinfo {pages} {075002} (\bibinfo {year} {2009})}\BibitemShut {NoStop}%
\bibitem [{\citenamefont {Tanizawa}\ \emph {et~al.}(2016)\citenamefont {Tanizawa}, \citenamefont {Takahashi},\ and\ \citenamefont {Dixon}}]{Tanizawa2016}%
  \BibitemOpen
  \bibfield  {author} {\bibinfo {author} {\bibfnamefont {Y.}~\bibnamefont {Tanizawa}}, \bibinfo {author} {\bibfnamefont {R.}~\bibnamefont {Takahashi}},\ and\ \bibinfo {author} {\bibfnamefont {A.~R.}\ \bibnamefont {Dixon}},\ }in\ \href {https://doi.org/10.1109/ICUFN.2016.7537018} {\emph {\bibinfo {booktitle} {2016 Eighth International Conference on Ubiquitous and Future Networks (ICUFN)}}}\ (\bibinfo {year} {2016})\ pp.\ \bibinfo {pages} {208--214}\BibitemShut {NoStop}%
\bibitem [{\citenamefont {Caleffi}(2017)}]{Caleffi2017}%
  \BibitemOpen
  \bibfield  {author} {\bibinfo {author} {\bibfnamefont {M.}~\bibnamefont {Caleffi}},\ }\href {https://doi.org/10.1109/ACCESS.2017.2763325} {\bibfield  {journal} {\bibinfo  {journal} {IEEE Access}\ }\textbf {\bibinfo {volume} {5}},\ \bibinfo {pages} {22299} (\bibinfo {year} {2017})}\BibitemShut {NoStop}%
\bibitem [{\citenamefont {Chao~Yang}\ and\ \citenamefont {Su}(2017)}]{Su2017}%
  \BibitemOpen
  \bibfield  {author} {\bibinfo {author} {\bibfnamefont {H.~Z.}\ \bibnamefont {Chao~Yang}}\ and\ \bibinfo {author} {\bibfnamefont {J.}~\bibnamefont {Su}},\ }\href {https://doi.org/10.1080/09500340.2017.1360956} {\bibfield  {journal} {\bibinfo  {journal} {Journal of Modern Optics}\ }\textbf {\bibinfo {volume} {64}},\ \bibinfo {pages} {2350} (\bibinfo {year} {2017})},\ \Eprint {https://arxiv.org/abs/https://doi.org/10.1080/09500340.2017.1360956} {https://doi.org/10.1080/09500340.2017.1360956} \BibitemShut {NoStop}%
\bibitem [{\citenamefont {Ma}\ \emph {et~al.}(2017)\citenamefont {Ma}, \citenamefont {Guo},\ and\ \citenamefont {Su}}]{ma2017multiple}%
  \BibitemOpen
  \bibfield  {author} {\bibinfo {author} {\bibfnamefont {C.}~\bibnamefont {Ma}}, \bibinfo {author} {\bibfnamefont {Y.}~\bibnamefont {Guo}},\ and\ \bibinfo {author} {\bibfnamefont {J.}~\bibnamefont {Su}},\ }in\ \href@noop {} {\emph {\bibinfo {booktitle} {2017 IEEE 2nd Advanced Information Technology, Electronic and Automation Control Conference (IAEAC)}}}\ (\bibinfo {organization} {IEEE},\ \bibinfo {year} {2017})\ pp.\ \bibinfo {pages} {2513--2517}\BibitemShut {NoStop}%
\bibitem [{\citenamefont {Tysowski}\ \emph {et~al.}(2018)\citenamefont {Tysowski}, \citenamefont {Ling}, \citenamefont {L\"utkenhaus},\ and\ \citenamefont {Mosca}}]{Mosca2018}%
  \BibitemOpen
  \bibfield  {author} {\bibinfo {author} {\bibfnamefont {P.~K.}\ \bibnamefont {Tysowski}}, \bibinfo {author} {\bibfnamefont {X.}~\bibnamefont {Ling}}, \bibinfo {author} {\bibfnamefont {N.}~\bibnamefont {L\"utkenhaus}},\ and\ \bibinfo {author} {\bibfnamefont {M.}~\bibnamefont {Mosca}},\ }\href {https://doi.org/10.1088/2058-9565/aa9a5d} {\bibfield  {journal} {\bibinfo  {journal} {Quantum Science and Technology}\ }\textbf {\bibinfo {volume} {3}},\ \bibinfo {pages} {024001} (\bibinfo {year} {2018})}\BibitemShut {NoStop}%
\bibitem [{\citenamefont {Mehic}\ \emph {et~al.}(2020)\citenamefont {Mehic}, \citenamefont {Niemiec}, \citenamefont {Rass}, \citenamefont {Ma}, \citenamefont {Peev}, \citenamefont {Aguado}, \citenamefont {Martin}, \citenamefont {Schauer}, \citenamefont {Poppe}, \citenamefont {Pacher},\ and\ \citenamefont {Voznak}}]{Mehic2020}%
  \BibitemOpen
  \bibfield  {author} {\bibinfo {author} {\bibfnamefont {M.}~\bibnamefont {Mehic}}, \bibinfo {author} {\bibfnamefont {M.}~\bibnamefont {Niemiec}}, \bibinfo {author} {\bibfnamefont {S.}~\bibnamefont {Rass}}, \bibinfo {author} {\bibfnamefont {J.}~\bibnamefont {Ma}}, \bibinfo {author} {\bibfnamefont {M.}~\bibnamefont {Peev}}, \bibinfo {author} {\bibfnamefont {A.}~\bibnamefont {Aguado}}, \bibinfo {author} {\bibfnamefont {V.}~\bibnamefont {Martin}}, \bibinfo {author} {\bibfnamefont {S.}~\bibnamefont {Schauer}}, \bibinfo {author} {\bibfnamefont {A.}~\bibnamefont {Poppe}}, \bibinfo {author} {\bibfnamefont {C.}~\bibnamefont {Pacher}},\ and\ \bibinfo {author} {\bibfnamefont {M.}~\bibnamefont {Voznak}},\ }\href {https://doi.org/10.1145/3402192} {\bibfield  {journal} {\bibinfo  {journal} {ACM Comput. Surv.}\ }\textbf {\bibinfo {volume} {53}},\ \bibinfo {pages} {1} (\bibinfo {year} {2020})}\BibitemShut {NoStop}%
\bibitem [{\citenamefont {Amer}\ \emph {et~al.}(2020)\citenamefont {Amer}, \citenamefont {Krawec},\ and\ \citenamefont {Wang}}]{Amer2020}%
  \BibitemOpen
  \bibfield  {author} {\bibinfo {author} {\bibfnamefont {O.}~\bibnamefont {Amer}}, \bibinfo {author} {\bibfnamefont {W.~O.}\ \bibnamefont {Krawec}},\ and\ \bibinfo {author} {\bibfnamefont {B.}~\bibnamefont {Wang}},\ }in\ \href {https://doi.org/10.1109/QCE49297.2020.00027} {\emph {\bibinfo {booktitle} {2020 IEEE International Conference on Quantum Computing and Engineering (QCE)}}}\ (\bibinfo {year} {2020})\ pp.\ \bibinfo {pages} {137--147}\BibitemShut {NoStop}%
\bibitem [{\citenamefont {Chen}\ \emph {et~al.}(2022)\citenamefont {Chen}, \citenamefont {Zhang}, \citenamefont {Zhao}, \citenamefont {Yu},\ and\ \citenamefont {Liu}}]{Chen2022}%
  \BibitemOpen
  \bibfield  {author} {\bibinfo {author} {\bibfnamefont {L.}~\bibnamefont {Chen}}, \bibinfo {author} {\bibfnamefont {Z.}~\bibnamefont {Zhang}}, \bibinfo {author} {\bibfnamefont {M.}~\bibnamefont {Zhao}}, \bibinfo {author} {\bibfnamefont {K.}~\bibnamefont {Yu}},\ and\ \bibinfo {author} {\bibfnamefont {S.}~\bibnamefont {Liu}},\ }\bibfield  {journal} {\bibinfo  {journal} {Entropy}\ }\textbf {\bibinfo {volume} {24}},\ \href {https://doi.org/10.3390/e24111519} {10.3390/e24111519} (\bibinfo {year} {2022})\BibitemShut {NoStop}%
\bibitem [{\citenamefont {Chen}\ \emph {et~al.}(2023)\citenamefont {Chen}, \citenamefont {Chen}, \citenamefont {Chen},\ and\ \citenamefont {Zhao}}]{Chen2023}%
  \BibitemOpen
  \bibfield  {author} {\bibinfo {author} {\bibfnamefont {L.-Q.}\ \bibnamefont {Chen}}, \bibinfo {author} {\bibfnamefont {J.-Q.}\ \bibnamefont {Chen}}, \bibinfo {author} {\bibfnamefont {Q.-Y.}\ \bibnamefont {Chen}},\ and\ \bibinfo {author} {\bibfnamefont {Y.-L.}\ \bibnamefont {Zhao}},\ }\bibfield  {journal} {\bibinfo  {journal} {Quantum Information Processing}\ }\textbf {\bibinfo {volume} {22}},\ \href {https://doi.org/10.1007/s11128-022-03825-x} {10.1007/s11128-022-03825-x} (\bibinfo {year} {2023})\BibitemShut {NoStop}%
\bibitem [{\citenamefont {Bi}\ \emph {et~al.}(2023)\citenamefont {Bi}, \citenamefont {Miao},\ and\ \citenamefont {Di}}]{Bi2023}%
  \BibitemOpen
  \bibfield  {author} {\bibinfo {author} {\bibfnamefont {L.}~\bibnamefont {Bi}}, \bibinfo {author} {\bibfnamefont {M.}~\bibnamefont {Miao}},\ and\ \bibinfo {author} {\bibfnamefont {X.}~\bibnamefont {Di}},\ }\bibfield  {journal} {\bibinfo  {journal} {Applied Sciences}\ }\textbf {\bibinfo {volume} {13}},\ \href {https://doi.org/10.3390/app13158690} {10.3390/app13158690} (\bibinfo {year} {2023})\BibitemShut {NoStop}%
\bibitem [{\citenamefont {Dervisevic}\ \emph {et~al.}(2024)\citenamefont {Dervisevic}, \citenamefont {Tankovic}, \citenamefont {Fazel}, \citenamefont {Kompella}, \citenamefont {Fazio}, \citenamefont {Voznak},\ and\ \citenamefont {Mehic}}]{Dervisevic2024}%
  \BibitemOpen
  \bibfield  {author} {\bibinfo {author} {\bibfnamefont {E.}~\bibnamefont {Dervisevic}}, \bibinfo {author} {\bibfnamefont {A.}~\bibnamefont {Tankovic}}, \bibinfo {author} {\bibfnamefont {E.}~\bibnamefont {Fazel}}, \bibinfo {author} {\bibfnamefont {R.}~\bibnamefont {Kompella}}, \bibinfo {author} {\bibfnamefont {P.}~\bibnamefont {Fazio}}, \bibinfo {author} {\bibfnamefont {M.}~\bibnamefont {Voznak}},\ and\ \bibinfo {author} {\bibfnamefont {M.}~\bibnamefont {Mehic}},\ }\href@noop {} {\bibinfo {title} {{Quantum Key Distribution Networks -- Key Management: A Survey}}} (\bibinfo {year} {2024}),\ \Eprint {https://arxiv.org/abs/2408.04580} {arXiv:2408.04580} \BibitemShut {NoStop}%
\bibitem [{\citenamefont {Tang}\ \emph {et~al.}(2006)\citenamefont {Tang}, \citenamefont {Ma}, \citenamefont {Mink}, \citenamefont {Nakassis}, \citenamefont {Xu}, \citenamefont {Hershman}, \citenamefont {Bienfang}, \citenamefont {Su}, \citenamefont {Boisvert}, \citenamefont {Clark},\ and\ \citenamefont {Williams}}]{Tang2006}%
  \BibitemOpen
  \bibfield  {author} {\bibinfo {author} {\bibfnamefont {X.}~\bibnamefont {Tang}}, \bibinfo {author} {\bibfnamefont {L.}~\bibnamefont {Ma}}, \bibinfo {author} {\bibfnamefont {A.}~\bibnamefont {Mink}}, \bibinfo {author} {\bibfnamefont {A.}~\bibnamefont {Nakassis}}, \bibinfo {author} {\bibfnamefont {H.}~\bibnamefont {Xu}}, \bibinfo {author} {\bibfnamefont {B.}~\bibnamefont {Hershman}}, \bibinfo {author} {\bibfnamefont {J.}~\bibnamefont {Bienfang}}, \bibinfo {author} {\bibfnamefont {D.}~\bibnamefont {Su}}, \bibinfo {author} {\bibfnamefont {R.~F.}\ \bibnamefont {Boisvert}}, \bibinfo {author} {\bibfnamefont {C.}~\bibnamefont {Clark}},\ and\ \bibinfo {author} {\bibfnamefont {C.}~\bibnamefont {Williams}},\ }in\ \href {https://doi.org/10.1117/12.679589} {\emph {\bibinfo {booktitle} {Quantum Communications and Quantum Imaging IV}}},\ Vol.\ \bibinfo {volume} {6305},\ \bibinfo {editor} {edited by\ \bibinfo {editor} {\bibfnamefont {R.~E.}\ \bibnamefont {Meyers}}, \bibinfo {editor} {\bibfnamefont {Y.}~\bibnamefont
  {Shih}},\ and\ \bibinfo {editor} {\bibfnamefont {K.~S.}\ \bibnamefont {Deacon}}},\ \bibinfo {organization} {International Society for Optics and Photonics}\ (\bibinfo  {publisher} {SPIE},\ \bibinfo {year} {2006})\ p.\ \bibinfo {pages} {630506}\BibitemShut {NoStop}%
\bibitem [{\citenamefont {Tayduganov}\ \emph {et~al.}(2021)\citenamefont {Tayduganov}, \citenamefont {Rodimin}, \citenamefont {Kiktenko}, \citenamefont {Kurochkin}, \citenamefont {Krivoshein}, \citenamefont {Khanenkov}, \citenamefont {Usova}, \citenamefont {Stefanenko}, \citenamefont {Kurochkin},\ and\ \citenamefont {Fedorov}}]{Tayduganov2021}%
  \BibitemOpen
  \bibfield  {author} {\bibinfo {author} {\bibfnamefont {A.}~\bibnamefont {Tayduganov}}, \bibinfo {author} {\bibfnamefont {V.}~\bibnamefont {Rodimin}}, \bibinfo {author} {\bibfnamefont {E.~O.}\ \bibnamefont {Kiktenko}}, \bibinfo {author} {\bibfnamefont {V.}~\bibnamefont {Kurochkin}}, \bibinfo {author} {\bibfnamefont {E.}~\bibnamefont {Krivoshein}}, \bibinfo {author} {\bibfnamefont {S.}~\bibnamefont {Khanenkov}}, \bibinfo {author} {\bibfnamefont {V.}~\bibnamefont {Usova}}, \bibinfo {author} {\bibfnamefont {L.}~\bibnamefont {Stefanenko}}, \bibinfo {author} {\bibfnamefont {Y.}~\bibnamefont {Kurochkin}},\ and\ \bibinfo {author} {\bibfnamefont {A.~K.}\ \bibnamefont {Fedorov}},\ }\href {https://doi.org/10.1364/OE.427804} {\bibfield  {journal} {\bibinfo  {journal} {Opt. Express}\ }\textbf {\bibinfo {volume} {29}},\ \bibinfo {pages} {24884} (\bibinfo {year} {2021})}\BibitemShut {NoStop}%
\bibitem [{\citenamefont {Salvail}\ \emph {et~al.}(2010)\citenamefont {Salvail}, \citenamefont {Peev}, \citenamefont {Diamanti}, \citenamefont {All{\'e}aume}, \citenamefont {L{\"u}tkenhaus},\ and\ \citenamefont {L{\"a}nger}}]{salvail2010security}%
  \BibitemOpen
  \bibfield  {author} {\bibinfo {author} {\bibfnamefont {L.}~\bibnamefont {Salvail}}, \bibinfo {author} {\bibfnamefont {M.}~\bibnamefont {Peev}}, \bibinfo {author} {\bibfnamefont {E.}~\bibnamefont {Diamanti}}, \bibinfo {author} {\bibfnamefont {R.}~\bibnamefont {All{\'e}aume}}, \bibinfo {author} {\bibfnamefont {N.}~\bibnamefont {L{\"u}tkenhaus}},\ and\ \bibinfo {author} {\bibfnamefont {T.}~\bibnamefont {L{\"a}nger}},\ }\href@noop {} {\bibfield  {journal} {\bibinfo  {journal} {Journal of Computer Security}\ }\textbf {\bibinfo {volume} {18}},\ \bibinfo {pages} {61} (\bibinfo {year} {2010})}\BibitemShut {NoStop}%
\bibitem [{\citenamefont {Zhou}\ \emph {et~al.}(2022)\citenamefont {Zhou}, \citenamefont {Lv}, \citenamefont {Huang},\ and\ \citenamefont {Ma}}]{zhou2022quantum}%
  \BibitemOpen
  \bibfield  {author} {\bibinfo {author} {\bibfnamefont {H.}~\bibnamefont {Zhou}}, \bibinfo {author} {\bibfnamefont {K.}~\bibnamefont {Lv}}, \bibinfo {author} {\bibfnamefont {L.}~\bibnamefont {Huang}},\ and\ \bibinfo {author} {\bibfnamefont {X.}~\bibnamefont {Ma}},\ }\href {https://doi.org/10.1109/TNET.2021.3136943} {\bibfield  {journal} {\bibinfo  {journal} {IEEE/ACM Transactions on Networking}\ }\textbf {\bibinfo {volume} {30}},\ \bibinfo {pages} {1328} (\bibinfo {year} {2022})}\BibitemShut {NoStop}%
\bibitem [{\citenamefont {Gaidash}\ \emph {et~al.}(2022)\citenamefont {Gaidash}, \citenamefont {Miroshnichenko},\ and\ \citenamefont {Kozubov}}]{gaidash2022quantum}%
  \BibitemOpen
  \bibfield  {author} {\bibinfo {author} {\bibfnamefont {A.}~\bibnamefont {Gaidash}}, \bibinfo {author} {\bibfnamefont {G.}~\bibnamefont {Miroshnichenko}},\ and\ \bibinfo {author} {\bibfnamefont {A.}~\bibnamefont {Kozubov}},\ }\href {https://doi.org/10.1364/JOCN.457492} {\bibfield  {journal} {\bibinfo  {journal} {Journal of Optical Communications and Networking}\ }\textbf {\bibinfo {volume} {14}},\ \bibinfo {pages} {934} (\bibinfo {year} {2022})}\BibitemShut {NoStop}%
\bibitem [{\citenamefont {Solomons}\ \emph {et~al.}(2022)\citenamefont {Solomons}, \citenamefont {Fletcher}, \citenamefont {Aktas}, \citenamefont {Venkatachalam}, \citenamefont {Wengerowsky}, \citenamefont {Lon{\v{c}}ari{\'c}}, \citenamefont {Neumann}, \citenamefont {Liu}, \citenamefont {Samec}, \citenamefont {Stip{\v{c}}evi{\'c}} \emph {et~al.}}]{solomons2022scalable}%
  \BibitemOpen
  \bibfield  {author} {\bibinfo {author} {\bibfnamefont {N.~R.}\ \bibnamefont {Solomons}}, \bibinfo {author} {\bibfnamefont {A.~I.}\ \bibnamefont {Fletcher}}, \bibinfo {author} {\bibfnamefont {D.}~\bibnamefont {Aktas}}, \bibinfo {author} {\bibfnamefont {N.}~\bibnamefont {Venkatachalam}}, \bibinfo {author} {\bibfnamefont {S.}~\bibnamefont {Wengerowsky}}, \bibinfo {author} {\bibfnamefont {M.}~\bibnamefont {Lon{\v{c}}ari{\'c}}}, \bibinfo {author} {\bibfnamefont {S.~P.}\ \bibnamefont {Neumann}}, \bibinfo {author} {\bibfnamefont {B.}~\bibnamefont {Liu}}, \bibinfo {author} {\bibfnamefont {{\v{Z}}.}~\bibnamefont {Samec}}, \bibinfo {author} {\bibfnamefont {M.}~\bibnamefont {Stip{\v{c}}evi{\'c}}}, \emph {et~al.},\ }\href {https://doi.org/10.1103/PRXQuantum.3.020311} {\bibfield  {journal} {\bibinfo  {journal} {PRX quantum}\ }\textbf {\bibinfo {volume} {3}},\ \bibinfo {pages} {020311} (\bibinfo {year} {2022})}\BibitemShut {NoStop}%
\bibitem [{\citenamefont {St{\k{e}}pniak}\ and\ \citenamefont {Mielczarek}(2023)}]{Stkepniak2023}%
  \BibitemOpen
  \bibfield  {author} {\bibinfo {author} {\bibfnamefont {M.}~\bibnamefont {St{\k{e}}pniak}}\ and\ \bibinfo {author} {\bibfnamefont {J.}~\bibnamefont {Mielczarek}},\ }\href@noop {} {\bibfield  {journal} {\bibinfo  {journal} {Journal of Information Security and Applications}\ }\textbf {\bibinfo {volume} {78}},\ \bibinfo {pages} {103581} (\bibinfo {year} {2023})}\BibitemShut {NoStop}%
\bibitem [{\citenamefont {Chen}\ \emph {et~al.}(2021)\citenamefont {Chen}, \citenamefont {Jiang}, \citenamefont {Tang}, \citenamefont {Zhou}, \citenamefont {Yuan}, \citenamefont {Zhou}, \citenamefont {Wang}, \citenamefont {Liu}, \citenamefont {Chen}, \citenamefont {Liu}, \citenamefont {Zhang}, \citenamefont {Cui}, \citenamefont {Liang}, \citenamefont {Li}, \citenamefont {Mao}, \citenamefont {Wang}, \citenamefont {Feng}, \citenamefont {Chen}, \citenamefont {Zhang}, \citenamefont {Li}, \citenamefont {Liu}, \citenamefont {Peng}, \citenamefont {Ma}, \citenamefont {Zhao},\ and\ \citenamefont {Pan}}]{Pan2021}%
  \BibitemOpen
  \bibfield  {author} {\bibinfo {author} {\bibfnamefont {T.-Y.}\ \bibnamefont {Chen}}, \bibinfo {author} {\bibfnamefont {X.}~\bibnamefont {Jiang}}, \bibinfo {author} {\bibfnamefont {S.-B.}\ \bibnamefont {Tang}}, \bibinfo {author} {\bibfnamefont {L.}~\bibnamefont {Zhou}}, \bibinfo {author} {\bibfnamefont {X.}~\bibnamefont {Yuan}}, \bibinfo {author} {\bibfnamefont {H.}~\bibnamefont {Zhou}}, \bibinfo {author} {\bibfnamefont {J.}~\bibnamefont {Wang}}, \bibinfo {author} {\bibfnamefont {Y.}~\bibnamefont {Liu}}, \bibinfo {author} {\bibfnamefont {L.-K.}\ \bibnamefont {Chen}}, \bibinfo {author} {\bibfnamefont {W.-Y.}\ \bibnamefont {Liu}}, \bibinfo {author} {\bibfnamefont {H.-F.}\ \bibnamefont {Zhang}}, \bibinfo {author} {\bibfnamefont {K.}~\bibnamefont {Cui}}, \bibinfo {author} {\bibfnamefont {H.}~\bibnamefont {Liang}}, \bibinfo {author} {\bibfnamefont {X.-G.}\ \bibnamefont {Li}}, \bibinfo {author} {\bibfnamefont {Y.}~\bibnamefont {Mao}}, \bibinfo {author} {\bibfnamefont {L.-J.}\ \bibnamefont {Wang}}, \bibinfo
  {author} {\bibfnamefont {S.-B.}\ \bibnamefont {Feng}}, \bibinfo {author} {\bibfnamefont {Q.}~\bibnamefont {Chen}}, \bibinfo {author} {\bibfnamefont {Q.}~\bibnamefont {Zhang}}, \bibinfo {author} {\bibfnamefont {L.}~\bibnamefont {Li}}, \bibinfo {author} {\bibfnamefont {N.-L.}\ \bibnamefont {Liu}}, \bibinfo {author} {\bibfnamefont {C.-Z.}\ \bibnamefont {Peng}}, \bibinfo {author} {\bibfnamefont {X.}~\bibnamefont {Ma}}, \bibinfo {author} {\bibfnamefont {Y.}~\bibnamefont {Zhao}},\ and\ \bibinfo {author} {\bibfnamefont {J.-W.}\ \bibnamefont {Pan}},\ }\href {https://doi.org/10.1038/s41534-021-00474-3} {\bibfield  {journal} {\bibinfo  {journal} {npj Quantum Information}\ }\textbf {\bibinfo {volume} {7}},\ \bibinfo {pages} {134} (\bibinfo {year} {2021})}\BibitemShut {NoStop}%
\bibitem [{\citenamefont {Kirsanov}\ \emph {et~al.}(2023)\citenamefont {Kirsanov}, \citenamefont {Pastushenko}, \citenamefont {Kodukhov}, \citenamefont {Yarovikov}, \citenamefont {Sagingalieva}, \citenamefont {Kronberg}, \citenamefont {Pflitsch},\ and\ \citenamefont {Vinokur}}]{Kirsanov2023}%
  \BibitemOpen
  \bibfield  {author} {\bibinfo {author} {\bibfnamefont {N.~S.}\ \bibnamefont {Kirsanov}}, \bibinfo {author} {\bibfnamefont {V.~A.}\ \bibnamefont {Pastushenko}}, \bibinfo {author} {\bibfnamefont {A.~D.}\ \bibnamefont {Kodukhov}}, \bibinfo {author} {\bibfnamefont {M.~V.}\ \bibnamefont {Yarovikov}}, \bibinfo {author} {\bibfnamefont {A.~B.}\ \bibnamefont {Sagingalieva}}, \bibinfo {author} {\bibfnamefont {D.~A.}\ \bibnamefont {Kronberg}}, \bibinfo {author} {\bibfnamefont {M.}~\bibnamefont {Pflitsch}},\ and\ \bibinfo {author} {\bibfnamefont {V.~M.}\ \bibnamefont {Vinokur}},\ }\href {https://doi.org/10.1038/s41598-023-35579-6} {\bibfield  {journal} {\bibinfo  {journal} {Scientific Reports}\ }\textbf {\bibinfo {volume} {13}},\ \bibinfo {pages} {8756} (\bibinfo {year} {2023})}\BibitemShut {NoStop}%
\bibitem [{\citenamefont {Kirsanov}\ \emph {et~al.}(2024)\citenamefont {Kirsanov}, \citenamefont {Pastushenko}, \citenamefont {Kodukhov}, \citenamefont {Aliev}, \citenamefont {Yarovikov}, \citenamefont {Strizhak}, \citenamefont {Zarubin}, \citenamefont {Smirnov}, \citenamefont {Pflitsch},\ and\ \citenamefont {Vinokur}}]{kirsanov2024loss}%
  \BibitemOpen
  \bibfield  {author} {\bibinfo {author} {\bibfnamefont {N.}~\bibnamefont {Kirsanov}}, \bibinfo {author} {\bibfnamefont {V.}~\bibnamefont {Pastushenko}}, \bibinfo {author} {\bibfnamefont {A.}~\bibnamefont {Kodukhov}}, \bibinfo {author} {\bibfnamefont {A.}~\bibnamefont {Aliev}}, \bibinfo {author} {\bibfnamefont {M.}~\bibnamefont {Yarovikov}}, \bibinfo {author} {\bibfnamefont {D.}~\bibnamefont {Strizhak}}, \bibinfo {author} {\bibfnamefont {I.}~\bibnamefont {Zarubin}}, \bibinfo {author} {\bibfnamefont {A.}~\bibnamefont {Smirnov}}, \bibinfo {author} {\bibfnamefont {M.}~\bibnamefont {Pflitsch}},\ and\ \bibinfo {author} {\bibfnamefont {V.}~\bibnamefont {Vinokur}},\ }\bibfield  {journal} {\bibinfo  {journal} {Entropy}\ }\textbf {\bibinfo {volume} {26}},\ \href {https://doi.org/10.3390/e26060437} {10.3390/e26060437} (\bibinfo {year} {2024})\BibitemShut {NoStop}%
\bibitem [{\citenamefont {Wang}(2005)}]{Wang2005}%
  \BibitemOpen
  \bibfield  {author} {\bibinfo {author} {\bibfnamefont {X.-B.}\ \bibnamefont {Wang}},\ }\bibfield  {journal} {\bibinfo  {journal} {Phys. Rev. Lett.}\ }\textbf {\bibinfo {volume} {94}},\ \href {https://doi.org/10.1103/physrevlett.94.230503} {10.1103/physrevlett.94.230503} (\bibinfo {year} {2005})\BibitemShut {NoStop}%
\bibitem [{\citenamefont {Ma}\ \emph {et~al.}(2005)\citenamefont {Ma}, \citenamefont {Qi}, \citenamefont {Zhao},\ and\ \citenamefont {Lo}}]{Ma2005}%
  \BibitemOpen
  \bibfield  {author} {\bibinfo {author} {\bibfnamefont {X.}~\bibnamefont {Ma}}, \bibinfo {author} {\bibfnamefont {B.}~\bibnamefont {Qi}}, \bibinfo {author} {\bibfnamefont {Y.}~\bibnamefont {Zhao}},\ and\ \bibinfo {author} {\bibfnamefont {H.-K.}\ \bibnamefont {Lo}},\ }\href {https://doi.org/10.1103/PhysRevA.72.012326} {\bibfield  {journal} {\bibinfo  {journal} {Phys. Rev. A}\ }\textbf {\bibinfo {volume} {72}},\ \bibinfo {pages} {012326} (\bibinfo {year} {2005})},\ \Eprint {https://arxiv.org/abs/quant-ph/0503005} {arXiv:quant-ph/0503005} \BibitemShut {NoStop}%
\bibitem [{\citenamefont {Zhang}\ \emph {et~al.}(2017)\citenamefont {Zhang}, \citenamefont {Zhao}, \citenamefont {Razavi},\ and\ \citenamefont {Ma}}]{Zhang2017}%
  \BibitemOpen
  \bibfield  {author} {\bibinfo {author} {\bibfnamefont {Z.}~\bibnamefont {Zhang}}, \bibinfo {author} {\bibfnamefont {Q.}~\bibnamefont {Zhao}}, \bibinfo {author} {\bibfnamefont {M.}~\bibnamefont {Razavi}},\ and\ \bibinfo {author} {\bibfnamefont {X.}~\bibnamefont {Ma}},\ }\href {https://doi.org/10.1103/PhysRevA.95.012333} {\bibfield  {journal} {\bibinfo  {journal} {Phys. Rev. A}\ }\textbf {\bibinfo {volume} {95}},\ \bibinfo {pages} {012333} (\bibinfo {year} {2017})}\BibitemShut {NoStop}%
\bibitem [{\citenamefont {Trushechkin}\ \emph {et~al.}(2017)\citenamefont {Trushechkin}, \citenamefont {Kiktenko},\ and\ \citenamefont {Fedorov}}]{Trushechkin2017}%
  \BibitemOpen
  \bibfield  {author} {\bibinfo {author} {\bibfnamefont {A.~S.}\ \bibnamefont {Trushechkin}}, \bibinfo {author} {\bibfnamefont {E.~O.}\ \bibnamefont {Kiktenko}},\ and\ \bibinfo {author} {\bibfnamefont {A.~K.}\ \bibnamefont {Fedorov}},\ }\href {https://doi.org/10.1103/PhysRevA.96.022316} {\bibfield  {journal} {\bibinfo  {journal} {Phys. Rev. A}\ }\textbf {\bibinfo {volume} {96}},\ \bibinfo {pages} {022316} (\bibinfo {year} {2017})}\BibitemShut {NoStop}%
\bibitem [{\citenamefont {Trushechkin}\ \emph {et~al.}(2021)\citenamefont {Trushechkin}, \citenamefont {Kiktenko}, \citenamefont {Kronberg},\ and\ \citenamefont {Fedorov}}]{Trushechkin2021}%
  \BibitemOpen
  \bibfield  {author} {\bibinfo {author} {\bibfnamefont {A.~S.}\ \bibnamefont {Trushechkin}}, \bibinfo {author} {\bibfnamefont {E.~O.}\ \bibnamefont {Kiktenko}}, \bibinfo {author} {\bibfnamefont {D.~A.}\ \bibnamefont {Kronberg}},\ and\ \bibinfo {author} {\bibfnamefont {A.~K.}\ \bibnamefont {Fedorov}},\ }\href {https://doi.org/10.3367/UFNe.2020.11.038882} {\bibfield  {journal} {\bibinfo  {journal} {Physics-Uspekhi}\ }\textbf {\bibinfo {volume} {64}},\ \bibinfo {pages} {88} (\bibinfo {year} {2021})}\BibitemShut {NoStop}%
\end{thebibliography}%

\end{document}